\documentclass[aps,pra,longbibliography,superscriptaddress,twocolumn,amsmath,amssymb,showpacs,floatfix]{revtex4-1}
\usepackage[utf8x]{inputenc}
\usepackage[english]{babel}
\usepackage[T1]{fontenc}
\usepackage{lmodern}
\usepackage{amssymb}
\usepackage{bbold}
\usepackage{amsmath}
\usepackage{amsthm}
\usepackage[pdftex]{graphicx}
\usepackage{epstopdf}
\usepackage{cancel}
\usepackage{lipsum}
\usepackage{dcolumn}
\usepackage{bm}
\usepackage{color}
\usepackage{relsize,dsfont,mathrsfs,empheq,verbatim,upgreek}
\usepackage{etoolbox}
\usepackage[caption = false]{subfig}
\usepackage{wrapfig}
\usepackage{datetime}
\usepackage{pgfplots}
\usepackage{times}
\usepackage{hyperref}
\hypersetup{%
   bookmarksopen=false,
   bookmarksnumbered=true,
   pdfnewwindow=true,
   unicode=false,
   colorlinks=true,%
   citecolor=blue,
   linkcolor=black,
   urlcolor=blue,
   filecolor=blue
   }  
\definecolor{darkblue}{rgb}{0,0,0.5}
\definecolor{grey25}{rgb}{0.25,0.25,0.25}
\definecolor{DarkOrange1}{rgb}{1.00,0.50,0.00}
\definecolor{lightblue}{rgb}{0,0,0.8}
\definecolor{royalblue}{rgb}{0.25,0.41,0.88}
\definecolor{lightcyan}{rgb}{0,0.8,0.8}
\definecolor{MediumVioletRed}{rgb}{0.78,0.08,0.52}
\definecolor{gold}{rgb}{1.00,0.84,0.00}
\definecolor{OrangeRed}{rgb}{1.00,0.27,0.00}
\definecolor{darkgreen}{rgb}{0,0.5,0}
\definecolor{viola}{rgb}{0.58,0,0.72}
\definecolor{ocra}{rgb}{0.673,0.276,0.05}
\definecolor{turchesevioletto}{rgb}{0.354,0.17,0.48}

\DeclareMathOperator{\Tr}{Tr}
\DeclareMathOperator{\Li}{Li}

\begin{document}
\newcommand{\freiburg}{Physikalisches Institut, Albert-Ludwigs-Universität Freiburg, Hermann-Herder-Stra{\ss}e 3, D-79104 Freiburg i.~Br., Federal Republic of Germany}
\newcommand{\frias}{Freiburg Institute for Advanced Studies, Albert-Ludwigs-Universit\"at Freiburg, Albertstr.~19, D-79104 Freiburg i.~Br., Federal Republic of Germany}
\newcommand{\parma}{Dipartimento di Scienze Matematiche, Fisiche e Informatiche, Universit\`a di Parma, Campus Universitario, Parco area delle Scienze n. 7/a, 43124 Parma, Italy}
\newcommand{\infn}{INFN, Sezione di Milano Bicocca, Gruppo Collegato di Parma, 43124 Parma, Italy}
\newcommand{\usal}{Departamento de F\'isica Fundamental, Universidad de Salamanca, E-37008 Salamanca, Spain}
\title{Non-interacting many-particle quantum transport between finite reservoirs}
\author{Giulio Amato}
\affiliation{\freiburg}
\affiliation{\parma}
\affiliation{\infn}
\author{Heinz-Peter Breuer}
\affiliation{\freiburg}
\affiliation{\frias}
\author{Sandro Wimberger}
\affiliation{\parma}
\affiliation{\infn}
\author{Alberto Rodríguez}
\affiliation{\freiburg}
\affiliation{\usal}
\author{Andreas Buchleitner}
\email[]{a.buchleitner@physik.uni-freiburg.de}
\affiliation{\freiburg}
\affiliation{\frias}
\begin{abstract}
We present a formalism to study many-particle quantum transport across a lattice locally connected to two finite, non-stationary (bosonic or fermionic) reservoirs, both of which are in a thermal state. 
We show that, for conserved total particle number, a system of nonlinear quantum-classical master equations describes the concurrent many-particle time evolution on the lattice and in the reservoirs.
The finiteness of the reservoirs makes a macroscopic current emerge, which decreases exponentially in time, and asymptotically drives the many-particle configuration into an equilibrium state where the particle flow ceases. 
We analytically derive the time scale of this equilibration process, and, furthermore, investigate the imprint of many-particle interferences on the transport process.
\end{abstract}
\maketitle
\section{Introduction}

The study of quantum transport across a confining potential landscape connected to leads has been a subject of intense research in the past decades 
\cite{Datta1995,Dit1998,Ram1998,Beenakker1991,Datta2005}, mostly with a focus on electronic transport in micro- and nano-systems.
This led to fundamental results, such as the Landauer-B\"uttiker formulas for current and conductance \cite{Lan1957,But1986,Datta1995,vanWees1988}
and the development of diverse theoretical methods, imported, e.g., from many-body theory
\cite{Mei1992,jauho1994,Datta1995,JinHEOM2008}. 

More recently, 
the realization of similar transport scenarios on different physical platforms, such as cavity QED systems \cite{Houck2012,Koch2013}, in optomechanics \cite{Marquardt2013} and with ultracold atoms in optical potentials \cite{Bra2012,Schneider2012,Stadler2012,Gat2012,Bar2013,Bra2013,Leonard2014,Eckel2014,Eckel2014a,Labouvie2015,CRyu2015,Krinner2017,Leb2018,Lebrat2019a,Corman2019}, 
has raised new 
theoretical questions. 
These new experiments allow for a fine control of the physical parameters and a good isolation from unwanted degrees of freedom.
In particular, for ultracold atoms, e.g., the interparticle interaction can be tuned via Feshbach resonances, and, interestingly, the quantum statistical nature of the carriers can be changed from fermionic to bosonic. Furthermore, in contrast to electronic transport through solid state samples, some of these experiments directly probe the non-stationary state of the reservoirs \cite{Krinner2017,Leb2018}.

Within the theory of open quantum systems \cite{Breuer2007}, quantum transport has been extensively studied perturbatively using a master equation approach, leading to 
several interesting results 
\cite{HaEs2006,EsHa2007,Wic2007,Pep2010,Pro2012,Zni2013,Asa2013,Ber2013,Iva2013,Kul2014,Kul2015,Kordas2015,Kul2016,Seg2016,Hofer2017,Kol2017,Kol2018,Kolovsky2019}. 
Nonetheless, all of these approaches rest on the assumption of stationary reservoirs during the evolution, and hence cannot account for situations where 
the non-trivial dynamics of the reservoirs establishes a final equilibrium condition, as observed in recent experiments \cite{Krinner2017,Leb2018}. 
Intuitively, it appears plausible that an initial imbalance of a given conserved physical quantity (such as, e.g., the total particle number) between two reservoirs drives the redistribution of that very quantity, mediated by an associated current which fades away as equidistribution is approached. 

Here, we present a refined treatment to describe the non-trivially coupled system and reservoir dynamics of non-interacting fermionic or bosonic particles, under the constraint of a conserved total particle number. The reservoirs are assumed to evolve as time dependent grand canonical thermal states, thermalizing on time scales much shorter than any of the dynamical time scales here of interest. This leads to a set of coupled nonlinear classical (for the reservoirs) and quantum (for the Hamiltonian system connecting the reservoirs) master equations which generalize previous treatments in the literature, which were either treating the baths and system dynamics as independent \cite{Bruderer2012,Schaller2014,Purk2017,deVega2018}, or forfeiting the coherent system dynamics, handling the transport channel via non coupled energy levels \cite{Nietner2014const,Schaller2014,Gallego-Marcos2014}. 

We benchmark our equations by studying transport of neutral atoms across a lattice, and highlight distinct dynamical regimes and phenomena, from an initially coherent evolution, over a metastable regime characterized by a non-vanishing current, towards a final equilibrium with vanishing current and fully balanced reservoir states.

The manuscript is organized as follows: In Sec.~\ref{sec:opensystemapproach}, we introduce the microscopic transport model and briefly review the standard open system technique to describe the system evolution  in the typical framework of stationary reservoirs.
A case is made for the adoption of the \textit{local} master equation, over the \textit{global} master equation, to appropriately tackle the transport problem at hand. 
In Sec.~\ref{sec:transportfinite}, 
we derive the set of coupled  master equations that describe the joined dynamics of reservoirs and system, while ensuring total particle number conservation. 
Section \ref{sec:numAnalysis} is devoted to a detailed ---analytical and numerical--- analysis of the time evolution, in terms of the single particle density matrix. 
First, in Sec.~\ref{sec:NumDynInfRes}, we review the features of transport between stationary reservoirs, and of the final non-equilibrium steady state. 
Then, in Sec.~\ref{subsec:finite}, 
we scrutinize the different dynamical transport regimes in the case of finite reservoirs, the emergence of a current-carrying metastable state and of a new equilibration time scale.
Finally, in Sec.~\ref{sec:tpdm}, by analyzing current and density fluctuations for fermions and bosons, we demonstrate how our formalism furthermore allows to unveil signatures of many-particle interference in a transport setup. 
\section{Quantum transport: open system approach}
\label{sec:opensystemapproach}
\subsection{The model}
We first establish the basic ingredients for an open system theory of quantum transport between infinite (bosonic or fermionic) and therefore stationary particle reservoirs. Consider a one dimensional lattice (hereafter also system, $S$) Hamiltonian
\begin{equation}
H_S = \sum_{i=1}^M \varepsilon_{S,i} a_i^{\dagger} a_i - \sum_{i \neq j}^M J_{ij} a_i^{\dagger} a_j ,
\label{eq:lattHamiltonian}
\end{equation}
with $\varepsilon_{S,i}$ the on-site energies, and $J_{ij} = J_{ji}$ the tunneling coupling strengths between adjacent sites $i$ and $j$ $( i \neq j )$. The lattice connects two \textit{stationary} reservoirs, left $(L)$ and right $(R)$, with reservoir Hamiltonians
\begin{equation}
H_{\epsilon} = \sum_{\kappa} \varepsilon_{\epsilon,{\kappa}} a^{\dagger}_{\epsilon,{\kappa}} a_{\epsilon,{\kappa}} ,
\label{eq:freeBathHam}
\end{equation}
$\epsilon\in\{L,R\}$, and $\kappa$ identifying the available reservoir modes. Let the system-reservoir interaction Hamiltonian be given by 
\begin{equation}
\begin{aligned}
H_{\textrm{int}} =& \sum_{\kappa} \nu^{L} (\kappa ) [ {a}^{\dagger}_{L,\kappa} {a}_{1} +  {a}_{L,\kappa} {a}_{1}^{\dagger} ] \\
 &+ \sum_{\kappa} \nu^{R} (\kappa ) [ {a}^{\dagger}_{R,\kappa} {a}_{M} +  {a}_{R,\kappa} {a}_{M}^{\dagger}], 
\end{aligned}
\label{eq:intHamLocal}
\end{equation}
with $\nu^{L} (\kappa )$ $[\nu^{R} (\kappa)]$ the coupling strength between the first (last) lattice site and reservoir mode $L$ $(R)$, $ \kappa $.  
All annihilation and creation operators considered satisfy canonical (anti-) commutation relations
for (fermionic) bosonic particles.

\begin{figure}
\centering
\includegraphics[width=.95\columnwidth]{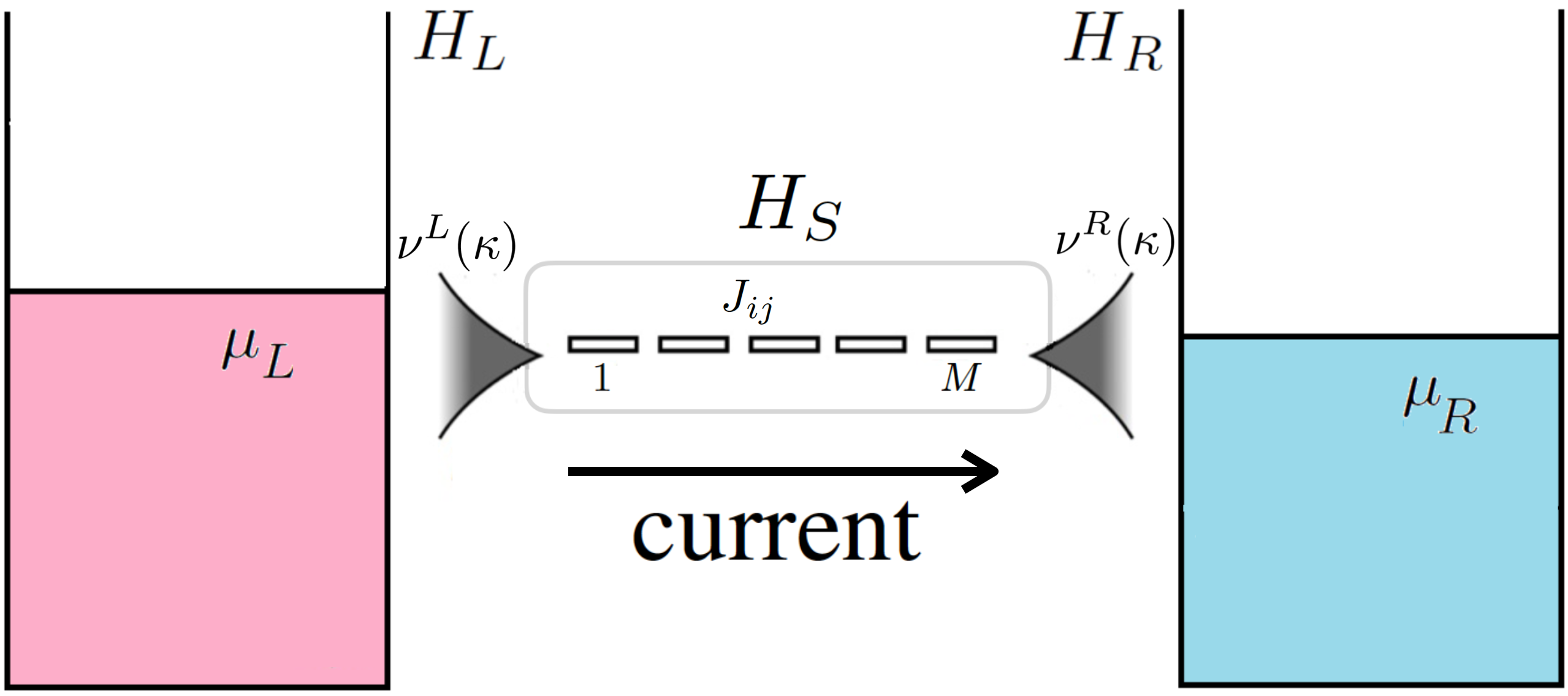}
\caption{
Microscopic quantum transport model of a system connecting two stationary reservoirs. The latter are described by grand canonical thermal states with fixed chemical potentials. The potential difference $\mu_L-\mu_R$ between the left ($L$) and right ($R$) reservoir induces the emergence of a non-interacting many particle current across a one dimensional, $ M$-site lattice with tunneling strength $J_{ij}$ between adjacent sites $i$ and $j$, until the reaching of a stationary condition. Only the terminal sites $ 1$ and $ M$ are coupled to the reservoirs $L$ and $R$, with coupling strengths $ \nu^L (\kappa) $ and $ \nu^R(\kappa ) $, respectively. The spectral structure of lattice and reservoirs, together with the respective couplings, are fixed by the associated Hamiltonians $ H_{L,S,R,\textrm{int}}$, see Eqs.~\eqref{eq:lattHamiltonian}-\eqref{eq:intHamLocal}.
} 
\label{fig:SetupInfRes}
\end{figure}

Let the reservoirs be initially prepared in their respective grand canonical thermal states
\begin{equation}
{\varrho}_{\epsilon} = \frac{1}{Z_{\epsilon}} e^{- \beta ( {H}_{\epsilon} - \mu_{\epsilon} {N}_{\epsilon} ) },
\label{eq:bathState}
\end{equation}
with $\hat{N}_{\epsilon} = \sum_{\kappa} {a}^{\dagger}_{\epsilon,\kappa} {a}_{\epsilon,\kappa} $ the number operator for reservoir $ \epsilon $ and $ Z_{\epsilon} = \Tr_{\epsilon }  \exp [ - \beta ( {H}_{\epsilon} - \mu_{\epsilon} \hat{N}_{\epsilon} )  ] $ the associated partition function. Note that this state is 
determined by the temperature $ T $ via $\beta\equiv (k_BT)^{-1}$ ---taken to be the same for both reservoirs---, and by the chemical potential $\mu_\epsilon$.
A schematic representation of the physical setting encoded by expressions \eqref{eq:lattHamiltonian}-\eqref{eq:bathState} is shown in Fig.~\ref{fig:SetupInfRes}.
\subsection{Master equation}
Under the assumption of a separation of time scales between the coherent many-particle dynamics on the lattice and the thermalization in the bath degree of freedom,
 standard projection operator techniques \cite{Breuer2007,Nakajima1958,Zwanzig1960} allow to write down an effective system evolution equation, in the form of a time-convolutionless master equation. At second order in the interaction Hamiltonian \eqref{eq:intHamLocal} (i.e., consistently, in the Born approximation), and in natural units $\hslash\equiv 1$, this equation assumes the Redfield form \cite{Breuer2007}
\begin{equation}
 \frac{d}{dt} \tilde{\rho}_{S} (t) = -  \int_{0}^t d \tau \Tr_E \left[ \tilde{H}_{\textrm{int}} (t), [ \tilde{H}_{\textrm{int}} ( t - \tau ) , \tilde{\rho}_{S} (t) \otimes \varrho_E ]\right] ,
\label{eq:preredfield}
\end{equation}
where $E$ lumps together both environments' ($L$+$R$)  degrees of freedom, i.e.,
$\varrho_E = \varrho_L \otimes \varrho_R$, and 
tilde quantities are given  in the interaction picture with respect to the free Hamiltonian $ H_0 = H_S + H_L + H_R $, with $H_\textrm{int}$ as in Eq.~\eqref{eq:intHamLocal} and $\rho_S$ the system state. 
To highlight the coherent and incoherent contributions to the dynamics, we write the system state in the Schr\"odinger picture, and hence Eq.~\eqref{eq:preredfield} reads
\begin{align}
 \frac{d}{dt} {\rho}_{S} (t) = & - i [ H_S , \rho_S (t) ] \notag \\
  &-  \int_{0}^t d \tau \Tr_E \left[ {H}_{\textrm{int}}, [ \tilde{H}_{\textrm{int}} (- \tau ) , {\rho}_{S} (t) \otimes \varrho_E ]\right].
\label{eq:redfield}
\end{align}

To ensure a well defined physical time evolution of the system degrees of freedom, we bring Eq.~\eqref{eq:redfield} into Lindblad form. 
Hereafter, we will adopt the \textit{local} master equation approach, in which 
the local coupling between system and reservoirs induces dissipative phenomena only on the extreme lattice sites. 
Note that, an alternative approach considered in the literature \cite{Kul2016,Hofer2017}, where the reservoirs couple globally to the 
system eigenmodes ---which may be spatially extended---
yields a suppression of the system coherences, due to the secular approximation involved \cite{Breuer2007}, which 
leads to vanishing site-to-site currents on the 
lattice \cite{Wic2007}.

The local master equation is obtained by elimination of the coherent coupling between edge and bulk sites in the incoherent part (the second line) of 
Eq.~\eqref{eq:redfield}. In fact, 
the Hamiltonian $\tilde{H}_{\textrm{int}} (- \tau )$
yields terms containing
$
\tilde{a}_1^{\dagger} (-\tau) = \exp \{ - i H_S \tau \}  \textrm{ } a_1^{\dagger} \exp \{ i H_S \tau \} 
$,
which, due to the form of the system Hamiltonian, can be rewritten as a time-dependent linear combination of the $a_j^{\dagger}$, with $ j \in \{ 1,...,M \}$. 
Assuming that 
\begin{equation}
    \begin{aligned}
    \tilde{a}_1^{\dagger} (-\tau) \approx & \textrm{ } \exp \{ - i \varepsilon_{S,1} a_1^{\dagger} a_1  \tau \}  \textrm{ } a_1^{\dagger} \exp \{ i \varepsilon_{S,1} a_1^{\dagger} a_1 \tau \} \\
    = & \textrm{ }  e^{ - i \varepsilon_{S,1}  \tau } a_1^{\dagger} ,
\end{aligned}  
\label{eq:dependencytimea1}
\end{equation}
which can be justified for $ \varepsilon_{S,1} \gtrsim J_{1j} $, for $ j \in \{1,...,M\} $ (see also Ref.~\cite{Hofer2017}), 
we approximate the interaction Hamiltonian, in the rotating frame generated by $ H_0 $, by
\begin{equation}
\tilde{H}_{\textrm{int}} (t) \approx \tilde{A}_{L}^{\dagger} (t) {a}_{1} + \tilde{A}_{L} (t) {a}_{L,\kappa}^{\dagger}  + \tilde{A}_{R}^{\dagger} (t) {a}_{M} + \tilde{A}_{R} (t) {a}_{M}^{\dagger},
\label{eq:intHamApp}
\end{equation}
with
\begin{equation}
\tilde{A}_{\epsilon} (t) = \sum_{\kappa} \nu^{\epsilon}( \kappa ) e^{ i ( \varepsilon_{S,1} - \varepsilon_{\epsilon,k} ) t } {a}_{\epsilon,\kappa} ,
\label{eq:exprA}
\end{equation}
and 
$\epsilon = \{ L,R \}$.
Using Eq.~\eqref{eq:intHamApp} in Eq.~\eqref{eq:redfield} and taking into account that $ \Tr_L [ \tilde{A}_L(t) \varrho_L ] = \Tr_L [ \tilde{A}^{\dagger}_L(t) \varrho_L ] = 0 $, one obtains, in the Markovian approximation and in the limit of 
continuum environmental modes \cite{Hofer2017},
\begin{equation}
\begin{aligned}
 \frac{{d}}{{d}t} {\rho}_{S} (t) =& \mathcal{L} \rho_S (t) \\
 \equiv &  - i [ H_S , \rho_S (t) ]  \\
  & +  \gamma_L n_{L} ( \varepsilon_{S,1} ) \mathcal{D} [ {a}_1^{\dagger} ] [ {\rho}_S (t) ] \\
  & + \gamma_L  [ 1 \pm n_{L} ( \varepsilon_{S,1} ) ] \mathcal{D} [ {a}_1 ] [ {\rho}_S (t) ] \\
  & + ( \{1,L \} \leftrightarrow \{ M,R \} ),
\end{aligned}
\label{eq:meLindblad}
\end{equation}
which exhibits the Gorini-Kossakowski-Sudarshan and Lindblad form \cite{Gorini1976,Lindblad1976a}. 
Here, the $+$ $(-)$ sign applies for bosons (fermions), and 
the notation $ ( \{1,L \} \leftrightarrow \{ M,R \} )$ 
stands for a repetition of the equation's third and fourth lines, with the indices $L$ and $1$ replaced by $ R$ and $M$, respectively. The dissipators in \eqref{eq:meLindblad} are given by 
\begin{equation}
\mathcal{D}[a] [ \rho_S (t) ] = a \rho_S (t)  a^{\dagger} - \frac{1}{2} \{ a^{\dagger} a ,  \rho_S (t)  \},
\label{eq:dissipator}
\end{equation}
and $ n_{\epsilon} ( \varepsilon )$ denotes the occupation number of energy level $\varepsilon$ in the
$\epsilon $ particle reservoir (in the continuum limit), according to the Bose-Einstein or Fermi-Dirac distribution, 
\begin{equation}
n_\epsilon (\varepsilon ) = \frac{1}{e^{\beta ( \varepsilon - \mu_\epsilon ) } \mp 1 }.
\label{eq:occN}
\end{equation}
The 
proportionality constant $\gamma_L$ in the particle gain and loss rates in Eq. \eqref{eq:meLindblad} is given by 
\begin{equation}
\gamma_L  = 2 \pi  [\nu^L (\varepsilon_{S,1} )]^2  D_L ( \varepsilon_{S,1} ),
\label{eq:rate}
\end{equation}
where $ \nu^L(\varepsilon)$ is the coupling constant from Eq.~\eqref{eq:intHamLocal} and
$D_L(\varepsilon)$ the left reservoir density of states, both evaluated at $ \varepsilon_{S,1} $ as implied by the Markov approximation (which keeps only resonant coupling terms between system and environment). The definition of $\gamma_R$ is strictly analogous.

We stress that the adopted master equation requires a separation between the correlation time scales in the reservoirs $ \tau_{\epsilon} $, with $ \epsilon \in \{ L,R \} $, and the system coherent time scales, i.e.,
\begin{equation}
 \varepsilon_{S,i } , J_{ij} \ll \tau_{\epsilon}^{-1}, 
\label{eq:sepTimeScales}
\end{equation}
and the weak-coupling or Born approximation, which is satisfied for
\begin{equation}
\gamma_{\epsilon} \ll \tau_{\epsilon}^{-1}
\label{eq:weakCoupling}
\end{equation}
as explained in Refs.~\cite{Breuer2007,vanKampen2007,Amato2019}. The bath correlation times $ \tau_{\epsilon}$, which can be computed after choosing the respective spectral densities via a Fourier transform \cite{Breuer2007}, can be engineered to be small at will, as the functional form of the coupling strengths $ \nu^{\epsilon} (\kappa ) $ need not be specified in our analysis. We emphasize that the 
 weak-coupling approximation does not enforce 
  any condition between $ \gamma_{\epsilon} $ and $ J_{ij} $. Our above derivation neither 
depends on the geometry of the system, nor 
on the specific values of on-site energies $ \varepsilon_{S,i} $ and tunnelling parameters $ J_{ij} $, and can be readily generalised to account for particles with internal degrees of freedom, think, e.g., of 
the spin transport setup described in Ref.~\cite{Lebrat2019}. 

As extensively analyzed in the literature  \cite{Wic2007,Pep2010,Pro2012,Zni2013,Asa2013,Ber2013,Iva2013,Levy2014,Kul2014,Kul2015,Kul2016,Seg2016,Hofer2017,Kol2017,Kol2018,DeChiara2018,Kolovsky2019}, stationary reservoirs drive the system into a unique \cite{Evans1977,Buca2012} non-equilibrium steady state (NESS). The latter is associated to a time-independent current flowing across the system, which is obtained as stationary solution of \eqref{eq:meLindblad}.

\section{Transport between finite reservoirs}
\label{sec:transportfinite}
\subsection{Time-dependent master equation}
\begin{figure}
\centering
\includegraphics[width=0.95\columnwidth]{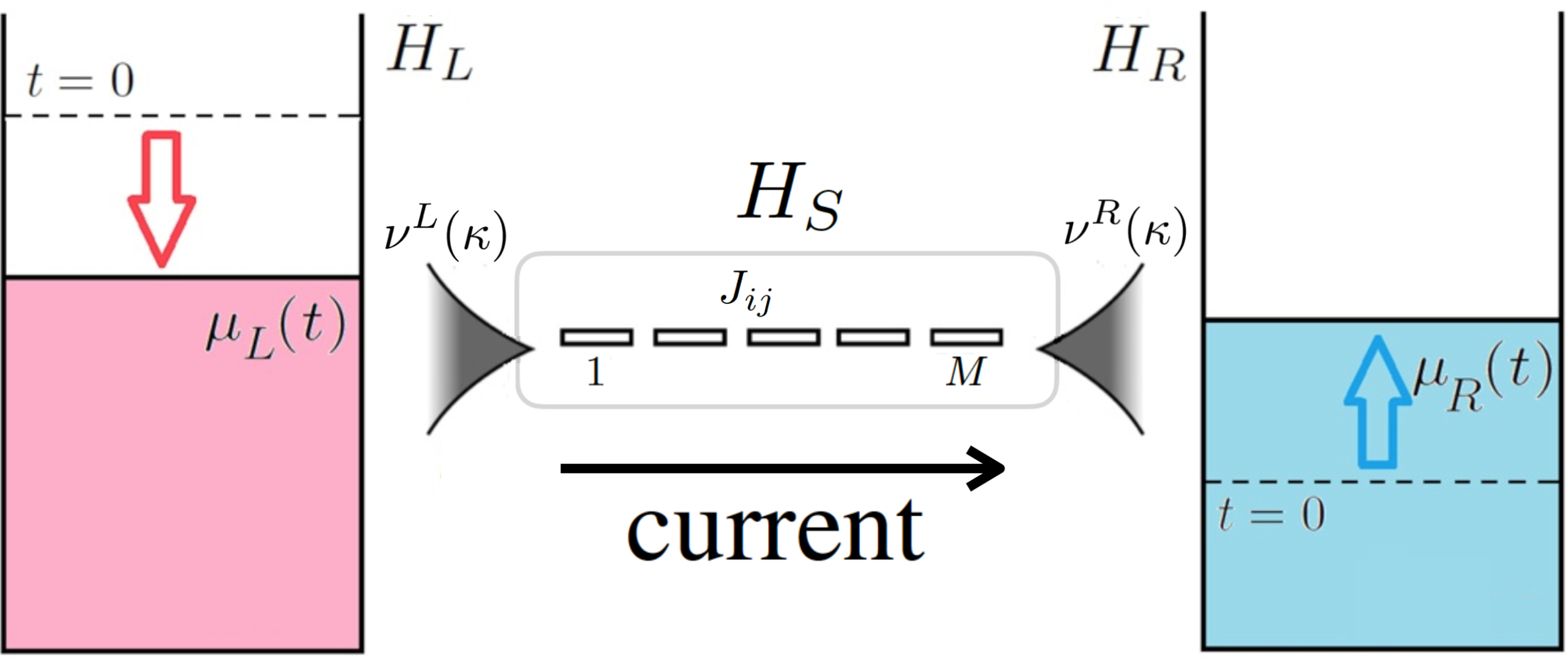}
\caption{Microscopic model with finite, non-stationary reservoirs, given an initial chemical potential bias $ [ \mu_L(t=0)-\mu_R(t=0) ] \neq 0 $. Model ingredients as in Fig.~\ref{fig:SetupInfRes}. The initial potential difference drives a non-equilibrium current across the lattice, which ceases as the reservoir populations equilibrate. The reservoirs are described as grand canonical thermal states with time-dependent chemical potentials.}
\label{fig:setupFinRes}
\end{figure}

Let us now replace the stationary reservoirs of the previous section by reservoirs which host a finite particle number, with an initial offset between the populations of $L$ and $R$, see Fig.~\ref{fig:setupFinRes}. The population difference corresponds to a chemical potential bias which drives a current across the lattice, which in turn mediates the equilibration of the population distribution over reservoirs and lattice, and must cease once the potential difference vanishes. In the following, we will assume both reservoirs to be described by thermal states (with time-dependent chemical potentials), equipped with the same, time-invariant spectral densities, and kept at identical temperature. Note that the latter assumption requires the same separation of time scales as already imposed for the case of infinite reservoirs above. 
For finite reservoirs, this implies, more specifically, that the particle number in each reservoir needs to be large as compared to the total particle number on the lattice, at all times.

Under these premises, we assume 
that the master equation describing the dynamics of the system 
still conserves the formal structure of a local master equation, 
accounting for a localized particle exchange with the reservoirs at the edges of the lattice. However, in order to cope with the 
evolution of the reservoirs, the rates $ \gamma_{\epsilon}^-(t)$ and $ \gamma_{\epsilon}^+(t)$ of particle loss and gain, respectively,  
are taken to be time dependent. Hence, the system dynamics is described by the explicitly time-dependent master equation
\begin{equation}
\begin{aligned}
 \frac{{d}}{{d}t} {\rho}_{S} (t) = &  \textrm{ } \mathcal{L}_t \rho_S(t) \\
 \equiv & - i  [ H_S , \rho_S (t) ] \\
 & + \gamma_L^+ (t) \mathcal{D} [ {a}_1^{\dagger} ] [ {\rho}_S (t) ] +  \gamma_L^- (t) \mathcal{D} [ {a}_1 ] [ {\rho}_S (t) ]  \\
 & + ( \{1,L \} \leftrightarrow \{ M,R \} ),
 \end{aligned}
\label{eq:meLindbladtime}
\end{equation}
where the rates $\gamma_\epsilon^{\pm} (t)$ remain to be determined. To guarantee a well defined 
evolution independently of the initial condition, i.e., to ensure complete positivity of the generated 
reduced dynamics \cite{Gorini1976,Lindblad1976a,BrLa2016}, we must require the rates $\gamma_\epsilon^{\pm} (t)$ to be non-negative, for all positive times.

From Eq.~\eqref{eq:meLindbladtime}, one can derive the evolution 
of the components of the single particle density matrix (SPDM), 
\begin{equation}
\sigma_{jk} (t) = \Tr_S [ a_j^{\dagger} a_k \rho_S (t) ],
\label{eq:defSPDM}
\end{equation}
via $\dot{\sigma}_{jk} (t) = \Tr_S [ a_j^{\dagger} a_k \mathcal{L}_t \rho_S (t) ]$. 
This will allow us to couple the reservoir and system time evolutions by the conservation of the total (finite) particle number in the tripartite system $ L + S + R $ (see further down).
For the sake of clarity, in the following we will restrict ourselves to the case where the system is a one-dimensional lattice with uniform nearest neighbour tunneling strengths $J_{ij}=J$ and identical on-site energies $\varepsilon_S$, $ \forall i,j $ in \eqref{eq:lattHamiltonian}. 
The evolution of the SPDM is then explicitly given by 
\begin{align}
\frac{d}{dt} {\sigma}_{jk} =&   i J [ \sigma_{j,k+1} + \sigma_{j,k-1} - \sigma_{j+1,k} - \sigma_{j-1,k} ]  \notag \\
& -  [ \gamma_L^- (t) \mathbf{\mp} \gamma_L^+ (t) ] \frac{ \delta_{1j} + \delta_{1k} }{2}  \sigma_{j k } + \delta_{1j} \delta_{1k} \gamma_L^+ (t) \notag \\
& +  ( \{1, L\} \leftrightarrow \{M,R\} ),
\label{eq:SPDM}
\end{align}
where the first line accounts for the Hamiltonian dynamics on the lattice,
and the remaining terms describe 
dissipative phenomena. The $-$ $(+)$ sign corresponds to the dynamics of bosons (fermions). 

Specifically, Eqs.~\eqref{eq:SPDM} 
determine the time evolution of the on-site occupation numbers 
\begin{equation}
n_l (t) = \sigma_{ll} (t), \qquad l \in \{1,\ldots,M\},
\end{equation}
the site-to-site currents (from site $l$ to site $l+1$)
\begin{equation}
j_{l,l+1 } (t) =  iJ  \left[ \sigma_{l+1,l} (t) - \sigma_{l,l+1}(t)\right], 
\label{eq:jcurrentdef}
\end{equation}
for $l \in \{1,\ldots,M-1\}$, and the long range coherences $|\sigma_{jk} (t)|$ between sites $j$ and $k\neq j,j\pm1$.

We emphasize that the differential equations \eqref{eq:SPDM} form 
a closed set. 
This feature stems from the quadratic form of the system Hamiltonian [Eq.~\eqref{eq:lattHamiltonian}], and does not hold in the presence of non-vanishing inter-particle interactions, leading, e.g., to quartic terms in creation and annihilation operators \cite{Kordas2015}.
\subsection{Conservation of the total particle number and coupled master equations}
The dissipative contribution to the time evolution of the on-site occupation number $n_1(t)$ can be extracted from Eq.~\eqref{eq:SPDM} to yield
the following equation,
\begin{equation}
\frac{d }{dt} n_1 (t ) \bigg|_\textrm{diss} = - \left[\gamma_L^- (t) \mp \gamma_L^+ (t) \right] n_1 (t ) + \gamma_L^+ (t) .
\label{eq:dissn1}
\end{equation}
The conservation of the total particle number enforces the requirement 
\begin{equation}
\frac{d }{dt} n_1 (t ) \bigg|_\textrm{diss}  = - \frac{d}{dt}  N_L (t),
\label{eq:consPartic}
\end{equation}
with the particle number $ N_L(t)$ in the left reservoir given by 
\begin{equation}
N_L (t) = \int_{E_0}^{\infty} d \varepsilon \, D (\varepsilon )  \textrm{ } n_L ( \varepsilon, t ),
\label{eq:defNL}
\end{equation}
in terms of the reservoir density of states $ D_L (\varepsilon )$, the minimum reservoir energy $ E_0 $, and the Bose-Einstein or Fermi-Dirac occupation number $ n_L ( \varepsilon, t )$ [Eq.~\eqref{eq:occN}], which depends on the time-dependent chemical potential $\mu_L(t)$. 

\begin{figure}
\centering
\includegraphics[width=0.85\columnwidth]{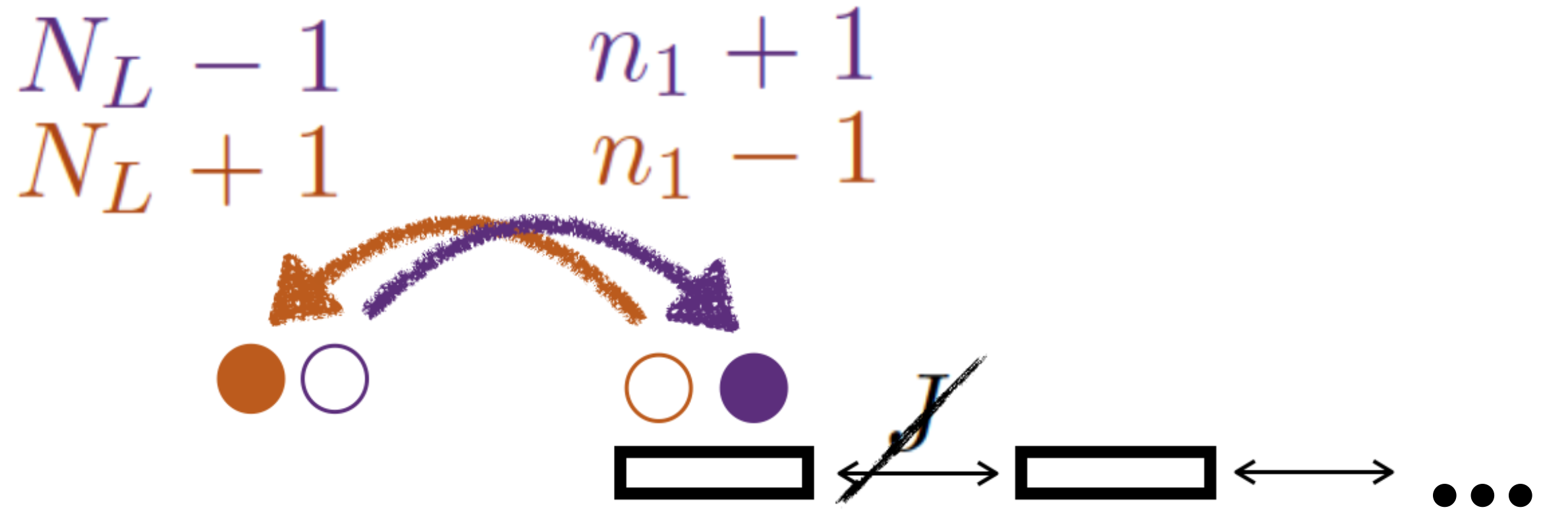}
\caption{Cartoon of the particle gain/loss relation [Eq.~\eqref{eq:consPartic}] between the left reservoir and the first lattice site when neglecting the contribution of coherent dynamics. Given $n_1 $ particles on the first site of the system, a particle can leave the system (empty circle), to be gained by the $ L $ reservoir (filled circle), leading to the transformations $ n_1 \rightarrow n_1 -1 $ and $ N_L \rightarrow N_L +1 $. Likewise, the loss of a particle in the left reservoir, $ N_L \rightarrow N_L - 1 $ (empty circle), leads to the gain of a particle (full circle) on the first site, $ n_1 \rightarrow n_1 +1 $. The differential version of this process is Eq.~\eqref{eq:consPartic}.}
\label{fig:n1nL}
\end{figure}

Equation \eqref{eq:consPartic} ensures that dissipative particle loss from the first site is compensated for by particle gain 
in the left reservoir, and vice versa, as depicted in Fig.~\ref{fig:n1nL}. 
The same relation holds for dissipative particle exchanges between site $M$ and the right reservoir $R$.

Recalling the Markovian approximation, 
which restricts the particle exchange between the lattice and both reservoirs to occur at the resonant energy $\varepsilon_S$, 
we propose the following classical rate equation 
for the evolution of the particle number in the left reservoir, 
\begin{equation}
\frac{d}{dt} N_L (t) = \gamma_L \left[ n_1 (t )- n_L (\varepsilon_{S}, t)\right],
\label{eq:NL}
\end{equation}
which states that its time derivative grows with the population of the first site but
decreases with the occupation number of the reservoir's resonant energy level. 
The parameter $\gamma_L$ 
would be determined by the underlying microscopic model [cf.~Eq.~\eqref{eq:rate}] 
and assumed to be time independent. Compatibility of 
Eqs.~\eqref{eq:dissn1}, \eqref{eq:consPartic} and \eqref{eq:NL} then imposes
\begin{equation}
\begin{aligned}
\gamma_L^+ (t) =  & \, \gamma_L n_L (\varepsilon_{S}, t ), \\
\gamma_L^- (t) =  & \, \gamma_L \left[ 1 \pm n_L (\varepsilon_{S}, t ) \right],
\end{aligned}
\label{eq:tDepRates}
\end{equation}
where $+$ ($-$) on the right hand side of the expression for $ \gamma_L^- (t) $ refers to the bosonic (fermionic) case.

Strictly analogous expressions are obtained for the transfer rates $ \gamma_R^{\pm} $ between site $ M$ and the right reservoir. We see that, as required, the rates $\gamma_{L,R}^{\pm}(t) $ are positive for all times, and that, moreover, they are obtained as straightforward generalizations of rates appearing in the master equation 
\eqref{eq:meLindblad} derived under the hypothesis of stationary reservoirs.

Altogether, the coupled evolution on the lattice and in the reservoirs is described by the following set of quantum-classical master equations:
\begin{subequations}
\begin{align}
\frac{{d}}{{d}t} {\rho}_{S} (t) = & \textrm{ } \mathcal{L}_t \rho_S(t), 
\label{eq:rhosevolution} \\ 
\frac{d}{dt} N_L (t) =  & \textrm{ } \gamma_L \left[  n_1 (t )- n_L (\varepsilon_{S}, t) \right], \label{eq:NLevolution}\\ 
\frac{d}{dt} N_R (t) =  & \textrm{ } \gamma_R \left[  n_M (t )- n_R (\varepsilon_{S}, t) \right] . \label{eq:NRevolution}
\end{align}
\label{eq:coupledEquations}
\end{subequations}

Note that both,
the reservoir particle number 
$N_L (t)$ [Eq.~\eqref{eq:defNL}] and the resonant energy level population 
$ n_L (\varepsilon_S, t) $ [Eq.~\eqref{eq:occN}], are determined by the chemical potential $\mu_L (t)$ (and likewise for the right reservoir). 
Hence, it is most convenient to write the reservoir evolution directly in terms of the chemical potential.
To this end, 
making use of Eqs.~\eqref{eq:defNL} and \eqref{eq:occN}, one can write 
\begin{equation}
 \frac{d}{dt}{N}_L (t) = f_L ( \mu_L (t) , \beta, E_0)  \textrm{ } \frac{d}{dt}\mu_L (t),
\end{equation}
where 
\begin{equation}
  f (\mu_L (t),\beta,E_0)=\int_{E_0}^{\infty} d\varepsilon \,D(\varepsilon)\, g(\mu_L(t), \beta, \varepsilon),
\label{eq:def_f}
\end{equation}
and 
\begin{equation}
 g(\mu_L(t), \beta, \varepsilon) = \frac{\beta e^{\beta \left[\varepsilon - \mu_L(t)\right]}}{\left(e^{\beta\left[\varepsilon - \mu_L(t)\right]} \mp 1 \right)^2 }
 \label{eq:def_g}
\end{equation}
is simply the derivative of the Bose-Einstein or Fermi-Dirac distribution with respect to the chemical potential. 
The function $f(\mu_L (t),\beta, E_0) $ is related to the underlying microscopic model of the reservoirs, since it implicitly depends on the reservoir density of states (cf.~Appendix \ref{sec:AppHarTrap}).

The equations for the coupled evolution of the system's SPDM [Eqs.~\eqref{eq:SPDM}] and the reservoirs can thus be cast as 
\begin{subequations}
\begin{align}
\frac{d}{dt} {\sigma}_{jk} = & i J \left[ \sigma_{j,k+1} + \sigma_{j,k-1} - \sigma_{j+1,k} - \sigma_{j-1,k} \right] \notag\\ 
 & -\textrm{ } \gamma_L \frac{ \delta_{1j} + \delta_{1k} }{2}  \sigma_{j k } + \delta_{1j} \delta_{1k} \gamma_L n_L (\varepsilon_{S}, t) \notag\\
 &+  \textrm{ } ( \{1, L\} \leftrightarrow \{M,R\} ), \label{eq:coupledevolutionSPDM}\\
\frac{d}{dt} \mu_L (t) = &   \dfrac{\gamma_L}{f ( \mu_L (t) ,\beta,  E_0 )}  \left[  n_1 (t )- n_L (\varepsilon_{S}, t) \right], \label{eq:evolution_muL}\\
\frac{d}{dt} \mu_R (t) = &  \dfrac{\gamma_R}{f ( \mu_R (t) , \beta,  E_0)}   \left[  n_M (t )- n_R (\varepsilon_{S}, t) \right], \label{eq:evolution_muR}
\end{align}
\label{eq:bigSet}
\end{subequations}
where we recall that the populations $n_{L,R}(\varepsilon_S,t)$ are given by Eq.~\eqref{eq:occN}. Note that the equations for the SPDM look formally the same for bosons and fermions, and the only remaining difference is due to the reservoir particle distribution which enters the dynamics via the function $f$.

We stress once more that, since there is no many-particle interaction within the lattice, Eqs.~\eqref{eq:bigSet} form 
a closed set of differential equations for the variables $\{\sigma_{jk}(t),\mu_L(t),\mu_R(t)\}$. Furthermore, it must be emphasized that the coupled system-reservoir evolution is governed by a set of equations, which is nonlinear in the chemical potentials $ \mu_{L,R} $.

Eqs.~\eqref{eq:coupledEquations} ---and consequently Eqs.\eqref{eq:bigSet}--- require a separation of the system/bath time scales [Eq.~\eqref{eq:sepTimeScales}], weak-coupling [Eq.~\eqref{eq:weakCoupling}], and a large number of particles in the reservoirs during the entire time-evolution, in order to justify the statistical description of the baths, i.e.,  through evolving grand-canonical thermal states. 
Note that in principle (numerically) exact treatments are available to model fermionic transport across a point contact, for infinite \cite{Schmidt2008} and finite size \cite{Kulkrani2013} reservoirs. These approaches would then need to be generalised to account for the equilibration dynamics between the reservoirs and the multi level junction here under consideration, for large particle numbers.

Finally, note that the 
time independent master equation \eqref{eq:meLindblad}, associated with stationary reservoirs, is retrieved from Eqs.~\eqref{eq:coupledEquations}, if we remove the time dependence of the reservoir occupation numbers in Eq.~\eqref{eq:rhosevolution}, and discard Eqs.~\eqref{eq:NLevolution} and \eqref{eq:NRevolution}.

\section{Dynamical regimes of quantum transport}
\label{sec:numAnalysis}
With the above results at hand, we can now proceed towards a detailed analysis of the coupled system-reservoir dynamics as generated by our time-dependent quantum-classical master equations. We numerically solve the nonlinear set of Eqs.~\eqref{eq:bigSet}, for a one-dimensional lattice with identical on-site energies $\varepsilon_S$ and nearest neighbour tunneling strength $ J$ (which is used as reference energy).
The reservoirs are described as 3D anisotropic harmonic traps (see Appendix \ref{sec:AppHarTrap}) with frequencies  $\omega_x = \omega_y =0.2J $, $\omega_z = 0.05 J$. We choose the effective lattice-reservoir coupling strengths to be $ \gamma_{L,R} = J/2 $.
The initial condition is chosen with no particles on the lattice, and with the reservoirs' temperature fixed at $\beta J=1 $.

\subsection{Stationary reservoirs}
\label{sec:NumDynInfRes}
To fully appreciate the fingerprint of a finite total particle number on the dynamics, let us first revisit the essential transport characteristics between infinite, stationary reservoirs. In this case 
the system dynamics are described by the master equation \eqref{eq:meLindblad}, and the state of the reservoirs is solely determined by the time-independent occupation numbers $n_{L,R}(\varepsilon_S)$
of the resonant energy levels.

\begin{figure}
\centering
 \includegraphics[width=0.9\columnwidth]{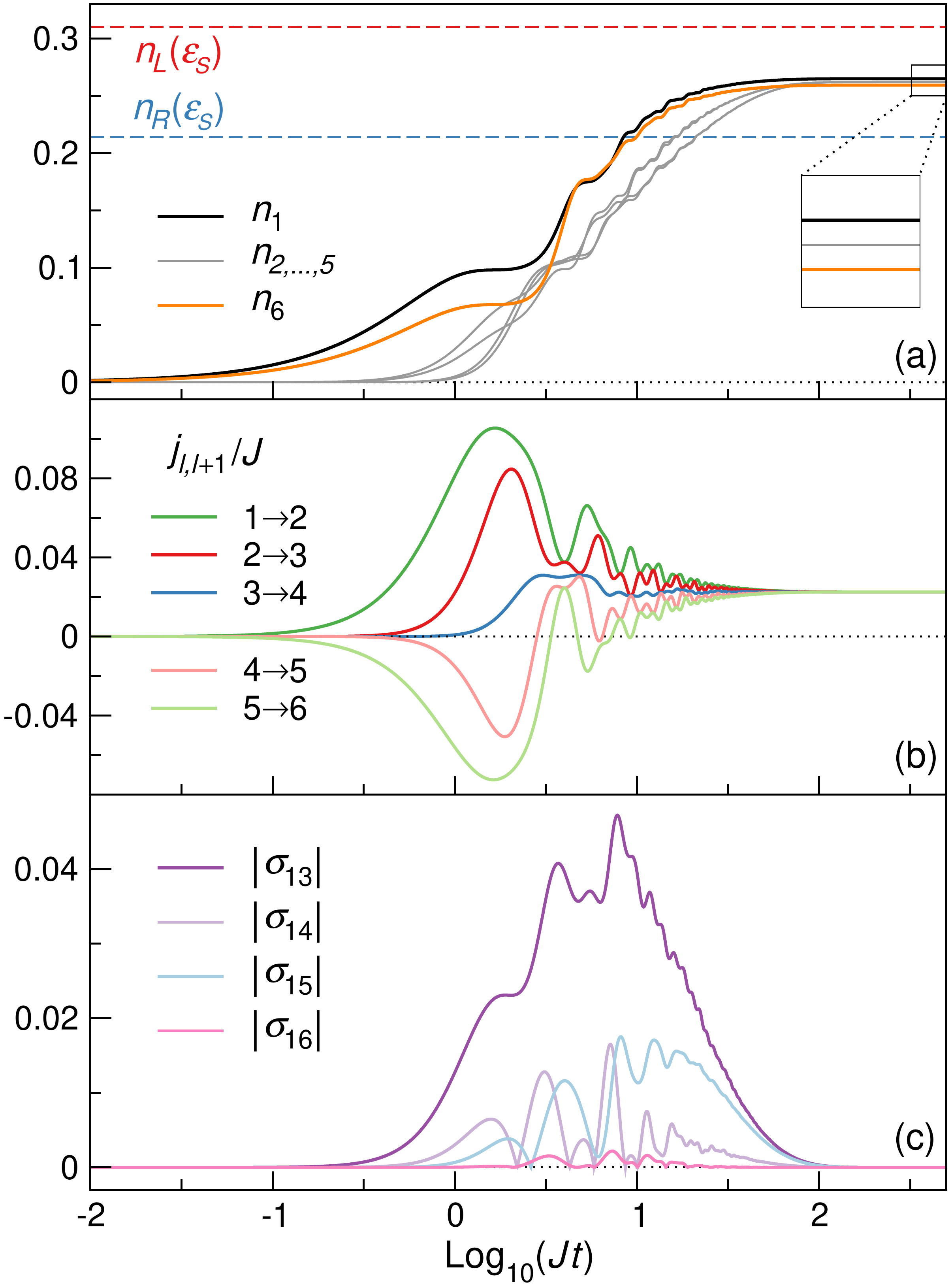}
\caption{Fermionic transport between stationary reservoirs. The different panels show the evolution of (a) populations of lattice sites and resonant reservoir modes, $n_l(t)$ and  $n_{L,R}(\varepsilon_S)$, respectively, (b) site-to-site currents $j_{l,l+1}(t)$, and (c) long-range coherences $\{\sigma_{1j}\}_{j=3,\ldots,6}$. 
The model parameters considered are $M=6$, $\varepsilon_S = 2J$, $ \gamma_L = \gamma_R = 0.5J $, $n_L(\varepsilon_S)=0.310$ and $ n_R(\varepsilon_S)=0.214$.}
\label{fig:exampleDynamics6}
\end{figure}

In Fig.~\ref{fig:exampleDynamics6}, we show the typical evolution of lattice site populations, site-to-site currents and long-range coherences. Starting from an empty lattice, the system undergoes an initial loading phase 
followed by a coherent dynamical transient phase, characterized by an oscillating (fluctuating) behaviour in time of the SPDM elements, as a consequence of single particle interference, for both fermionic and bosonic particles. 
For increasing time, the coherences $\sigma_{jk}(t)$ for $|k-j|>1$ decay to zero, as the system relaxes towards a non-equilibrium steady state (NESS) induced by the 
dissipative effects caused by the coupling to the reservoirs. 

\begin{figure}
\centering
 \includegraphics[width=0.9\columnwidth]{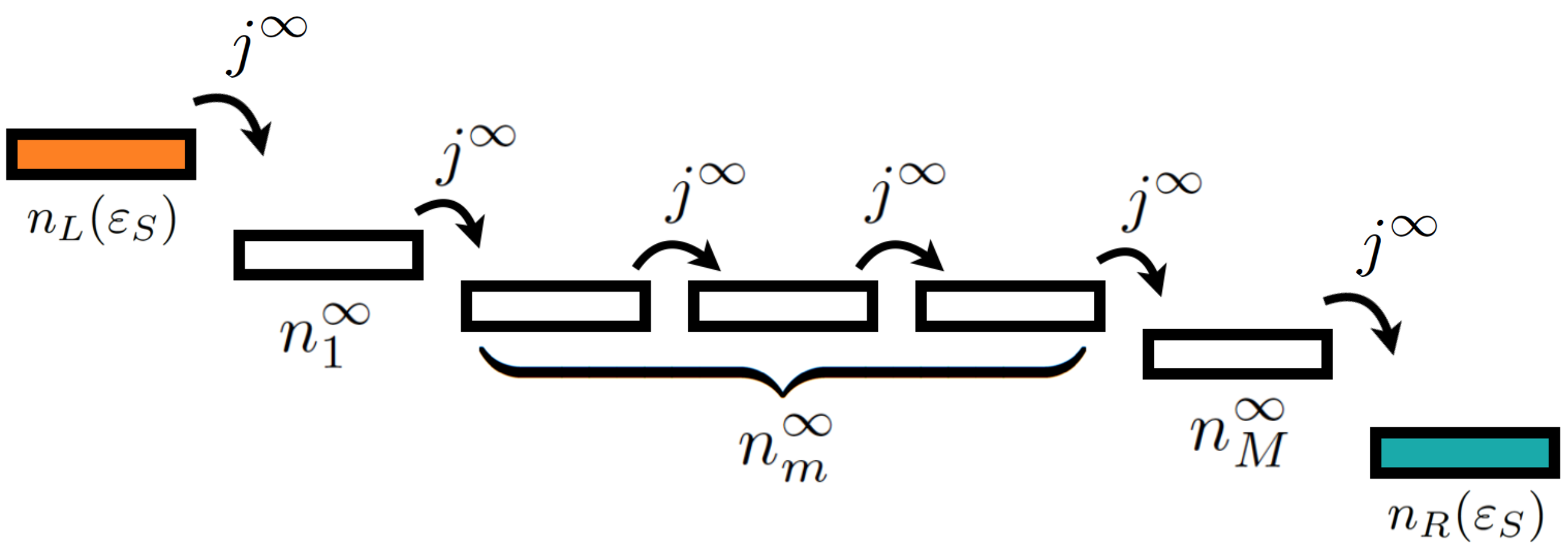}
\caption{
Representation of the non-equilibrium steady state established between infinite, stationary reservoirs. The asymptotic state is characterized by a uniform site-to-site current $j^\infty$ [Eq.~\eqref{eq:currentShortEquilibrium}] on the lattice, and a ladder-like structure of the on-site occupation numbers $n_l^\infty$
[Eqs.~\eqref{eq:populationsShortEquilibrium}].}
\label{fig:schemeInf}
\end{figure}

The NESS is defined by the fixed point
$\mathcal{L} \rho_S (t)=  0$ of Eq.~\eqref{eq:meLindblad}, which is reached in the limit $t\to\infty$. 
As already shown in Ref.~\cite{Asa2013}, and as also observed in Fig.~\ref{fig:exampleDynamics6}, the NESS is characterized by a uniform non-vanishing site-to-site current $j^\infty\equiv j_{l,l+1 } (t\to\infty)$,  
for all $l\in\{1,\ldots, M-1 \}$, 
\begin{equation}
j^{\infty} = \frac{ 4 \gamma_L \gamma_R J^2 }{ (4 J^2 + \gamma_L \gamma_R ) (\gamma_L + \gamma_R )} \Delta n, 
\label{eq:currentShortEquilibrium}
\end{equation}
a ladder like structure for the lattice populations $n_j^\infty\equiv n_j(t\to\infty)$ (depicted in Fig.~\ref{fig:schemeInf}), 
\begin{subequations}
\begin{align}
n_{1 }^{\infty}  = & \textrm{ } \bar{n} +  
\frac{ 4 (\gamma_L - \gamma_R) J^2 + \gamma_L \gamma_R^2 +  \gamma_L^2 \gamma_R }{ 2(4 J^2 + \gamma_L \gamma_R ) (\gamma_L + \gamma_R )} \Delta n,\\
n_{m }^{\infty} = & \textrm{ } \bar{n}  +  
\frac{ 4 (\gamma_L - \gamma_R) J^2 + \gamma_L \gamma_R^2 -  \gamma_L^2 \gamma_R }{ 2(4 J^2 + \gamma_L \gamma_R ) (\gamma_L + \gamma_R )} \Delta n,\\
n_{M }^{\infty} = & \textrm{ } \bar{n}  +  
\frac{ 4 (\gamma_L - \gamma_R) J^2 - \gamma_L \gamma_R^2 -  \gamma_L^2 \gamma_R }{ 2(4 J^2 + \gamma_L \gamma_R ) (\gamma_L + \gamma_R )} \Delta n,
\end{align}
\label{eq:populationsShortEquilibrium}
\end{subequations}
for $m\in\{2,\ldots, M-1\}$, 
and vanishing coherences 
$ \sigma_{jk}^\infty\equiv \sigma_{jk}(t\to\infty)=0$, for $|k-j|>1$.
All the asymptotic quantities above are given in terms of 
\begin{align}
\Delta n \equiv & \textrm{ } n_L ( \varepsilon_S  ) - n_R ( \varepsilon_S ), \label{eq:deltan}\\
\bar{ n} \equiv & \textrm{ } \frac{ n_L ( \varepsilon_S ) + n_R ( \varepsilon_S  )}{2}. \label{eq:barn}
\end{align}

\begin{figure}
\centering
\includegraphics[width=\columnwidth]{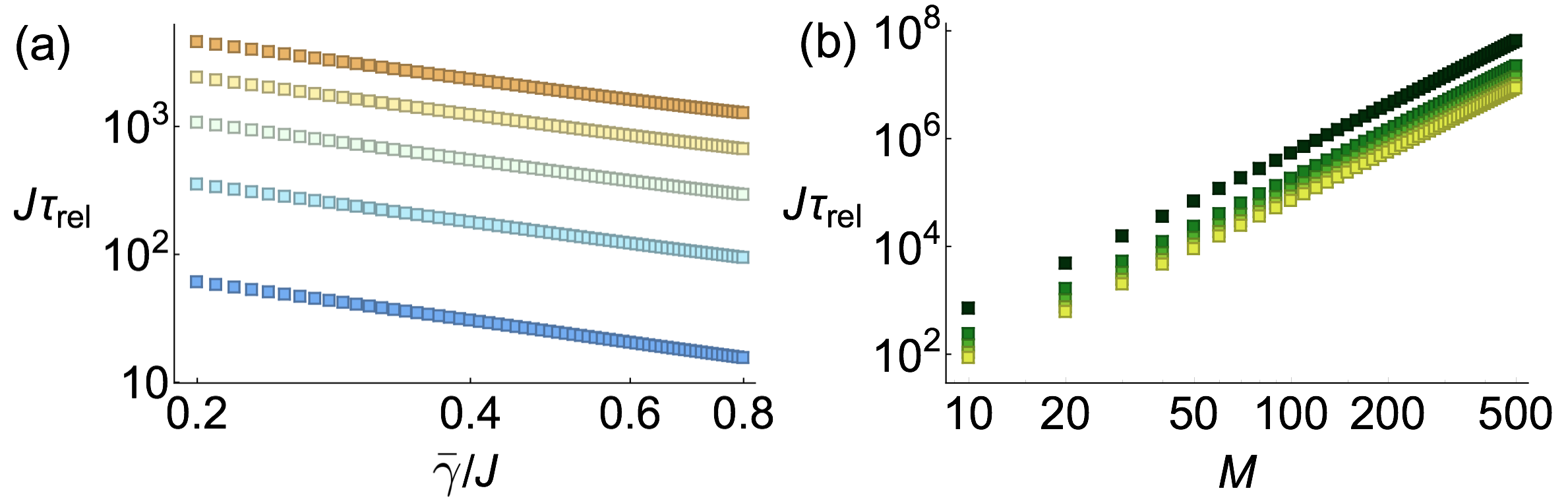}
\caption{Dependence of the 
relaxation time $\tau_{\textrm{rel}}$ [computed via Eq.~\eqref{eq:relaxTime} from the spectrum of the effective Hamiltonian \eqref{eq:effHam}] on the lattice length $M$ and the average coupling $\bar{\gamma} = (\gamma_L + \gamma_R)/2$: (a) Log-log plot of $ J \tau_{\textrm{rel}}$ on $\bar{\gamma}$, for lattice lengths $M=\{ 25,20,15,10,5\}$ (from top to bottom); (b) scaling of $\tau_{\textrm{rel}}$ with $M$, for $\bar{\gamma}=\{ 0.1, 0.3, 0.5,0.7,0.9 \}$ (from top to bottom).}
\label{fig:relaxTimeG}
\end{figure}

The dynamical approach towards the NESS is characterized by a relaxation time scale $\tau_{\textrm{rel}}$ which can be estimated by analyzing the spectral properties of the effective, non-Hermitian many-particle Hamiltonian on the lattice, 
\begin{equation}
H_{\textrm{eff}} = H_S - \frac{i}{2} \gamma_L a_1^{\dagger} a_1 - \frac{i}{2} \gamma_R a_M^{\dagger} a_M ,
\label{eq:effHam}
\end{equation}
which can be derived with the help of scattering theory  \cite{Gardiner1985,See2017}. This Hamiltonian is not sufficient to describe the full quantum transport problem 
but captures 
the decaying behaviour of the system's coherent dynamics. 
The imaginary parts of the effective Hamiltonian's complex eigenvalues, 
$E_k = \varepsilon_k - i\Gamma_k/2$, for $ k \in \{1,\ldots,M \}$, 
are nothing but the (exponential) decay rates of the associated eigenstates, with corresponding time scales $\tau_k\equiv( \Gamma_k)^{-1}$. 
For any non-trivial initial condition of the many-particle state, the dynamics on the lattice will therefore relax into the NESS on the time scale which fixes the life time of the longest-lived eigenstate of $H_\textrm{eff}$, and we therefore identify 
the relaxation time scale $\tau_{\textrm{rel}}$ of our open system with the slowest time scale associated to the eigenvectors of $H_\textrm{eff}$,
\begin{equation}
\tau_{\textrm{rel}} := \frac{ 1}{ \Gamma },
\label{eq:relaxTime}
\end{equation}
where $\Gamma\equiv \min \Gamma_k$.
The numerical analysis of $\tau_{\textrm{rel}}$ shown in Fig.~\ref{fig:relaxTimeG}, via diagonalization of $H_\textrm{eff}$, for variable 
$\gamma_L $,  $\gamma_R$ and number of sites $M$, 
reveals a dominant dependence of the form 
\begin{equation}
\tau_{\textrm{rel}} \propto \frac{ M^3 }{\bar{\gamma}},
\label{eq:relaxTimeDep}
\end{equation}
independently of the bosonic or fermionic character of the particles, with $\bar{\gamma} = (\gamma_L + \gamma_R)/2$. 
Equation \eqref{eq:relaxTimeDep}, also observed in Ref.~\cite{Kol2018}, is in agreement with previous studies 
that analyzed the eigenvalues of the Lindbladian superoperator in spin lattices \cite{Bi2014,Med2014,Zni2015}. 

\begin{figure}
\centering
\includegraphics[width=0.95\columnwidth]{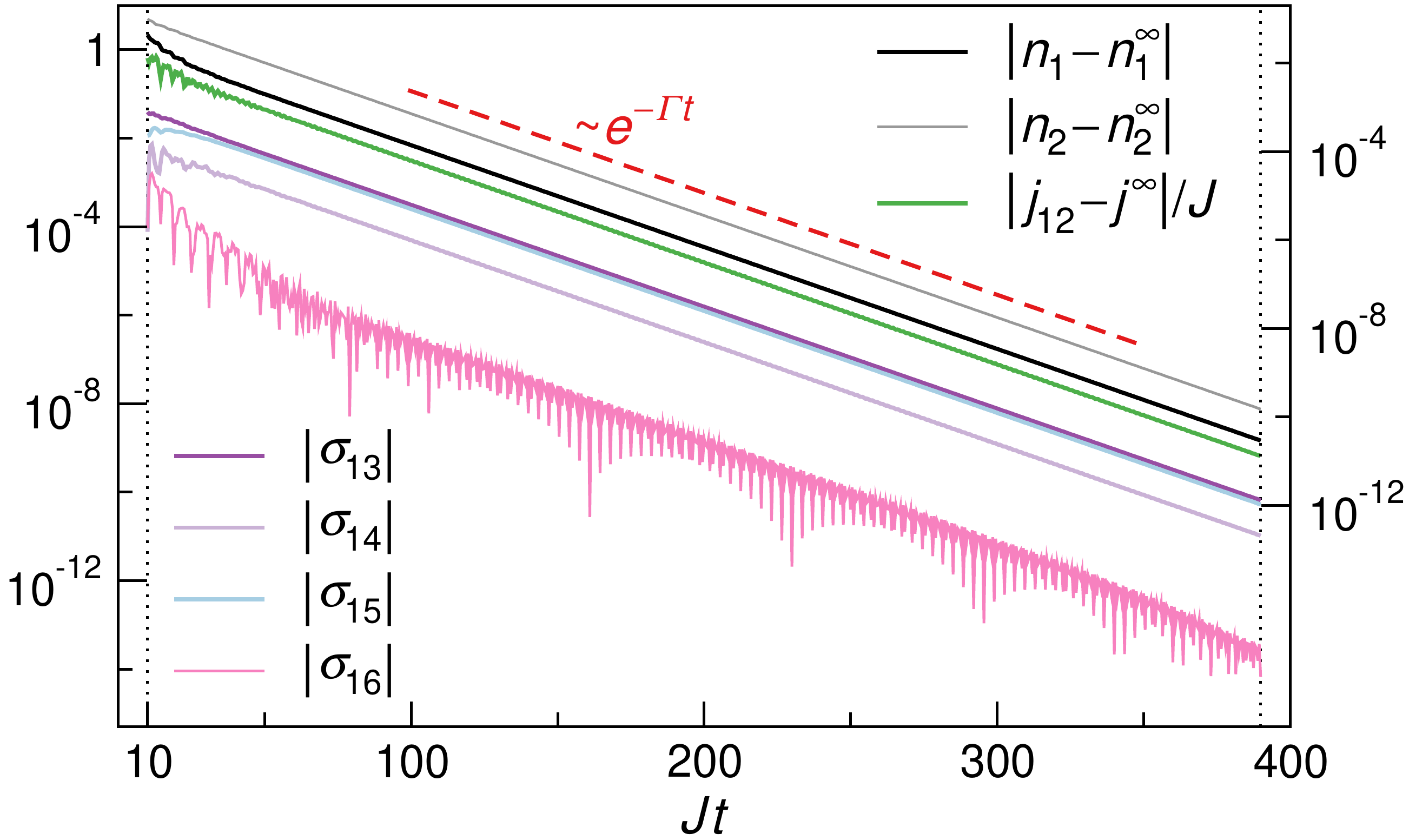}
\caption{Decay of fermionic on-site populations $n_l(t)$, current $j_{12}(t)$ (right ordinate), and long range coherences $\{ \sigma_{1j}\}_{j=3,\ldots,6}$ (left ordinate) with respect to their non-equilibrium steady state values. 
The dashed red line indicates exponential decay with rate $ \Gamma/J = 0.0530209 $ [which was extracted as the inverse life time of the longest lived eigenstate of the effective Hamiltonian \eqref{eq:effHam}] in perfect agreement with definition \eqref{eq:relaxTime}. Same parameter values as in Fig.~\ref{fig:exampleDynamics6}.
}
\label{fig:LongTimeDecayInfRes}
\end{figure}

In Fig.~\ref{fig:LongTimeDecayInfRes}, we show the relaxation dynamics of the SPDM elements, all of which exponentially converge into their respective NESS values with rate $\tau_{\textrm{rel}}^{-1}$. At sufficiently long times, quantum transport between stationary reservoirs is thus characterized by a single relevant time scale. 

\subsection{Finite reservoirs}
\label{subsec:finite}
For finite reservoirs, additional dynamical regimes amend the behaviour observed above, and a final (particle conserving) equilibrium state is naturally achieved.

The time evolution of the reservoir states is controlled by the constant reservoir temperature  (set to $\beta J=1$) and the time-dependent chemical potentials $\mu_{L,R}(t)$,  the latter being chosen at $t=0$  such as to define a finite potential bias between $L$ and $R$, and associated with finite occupation numbers  $N_{L,R}(0)\gg1$. Since initially no particles reside on the lattice, the total particle number $ N_0$ is given by
\begin{equation}
 N_0\equiv N_L(0)+N_R(0).
\end{equation} 

\begin{figure}
\centering
\includegraphics[width=.95\columnwidth]{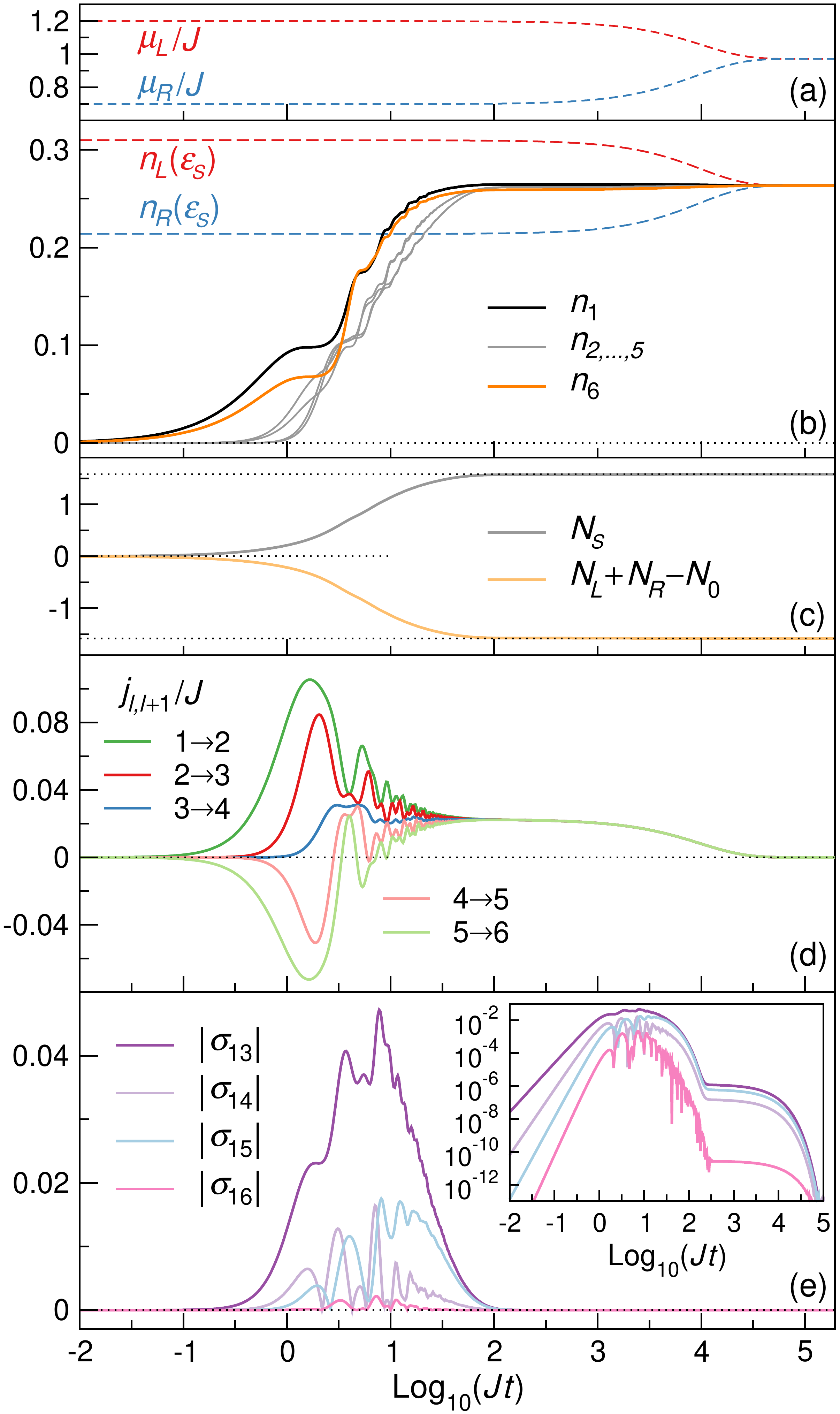}
\caption{Fermionic transport between finite (non-stationary) reservoirs obtained by numerical solution of Eqs.~\eqref{eq:bigSet}. The different panels show the evolution of (a) reservoir chemical potentials $\mu_{L,R}(t)$, (b) lattice site and resonant reservoir mode populations, $n_l(t)$ and $n_{L,R}(\varepsilon_S,t)$, respectively, (c) total particle number $N_S(t)$ on the lattice and particle deficit $N_L(t)+N_R(t)-N_0$ in the reservoirs, (d) site-to-site currents $j_{l,l+1}(t)$, and (e) long-range coherences $\{\sigma_{1j}\}_{j=3,\ldots,6}$. 
The parameters considered are $\mu_L(0)=1.2J$, $\mu_R(0)=0.7J$ [which correspond to $N_L(0)= 1276$, $N_R(0)=838$ and $n_L(\varepsilon_S,0)=0.310$, $n_R(\varepsilon_S,0)=0.214$], $M=6$, $\varepsilon_S=2J$, $\gamma_L=\gamma_R=0.5J$. This entails a final equilibrium state characterized by $\mu^\infty=0.972$, $n^\infty=0.263$, and $N^\infty=1056$.}
\label{fig:TransFerFinRes}
\end{figure}

A typical example of particle number conserving fermionic transport between non-stationary reservoirs is shown in Fig.~\ref{fig:TransFerFinRes}. The evolution can be divided into two major regimes: A short time coherent regime, where variations of the reservoir populations are negligible and the dynamics is equally well described by the time independent master equation \eqref{eq:meLindblad} (as for infinite reservoirs), and a long time regime, where the concurrent evolution of system and reservoir slowly converges into the final equilibrium condition.

\subsubsection{Short time coherent dynamics}
\begin{figure}
\centering
 \includegraphics[width=0.95\columnwidth]{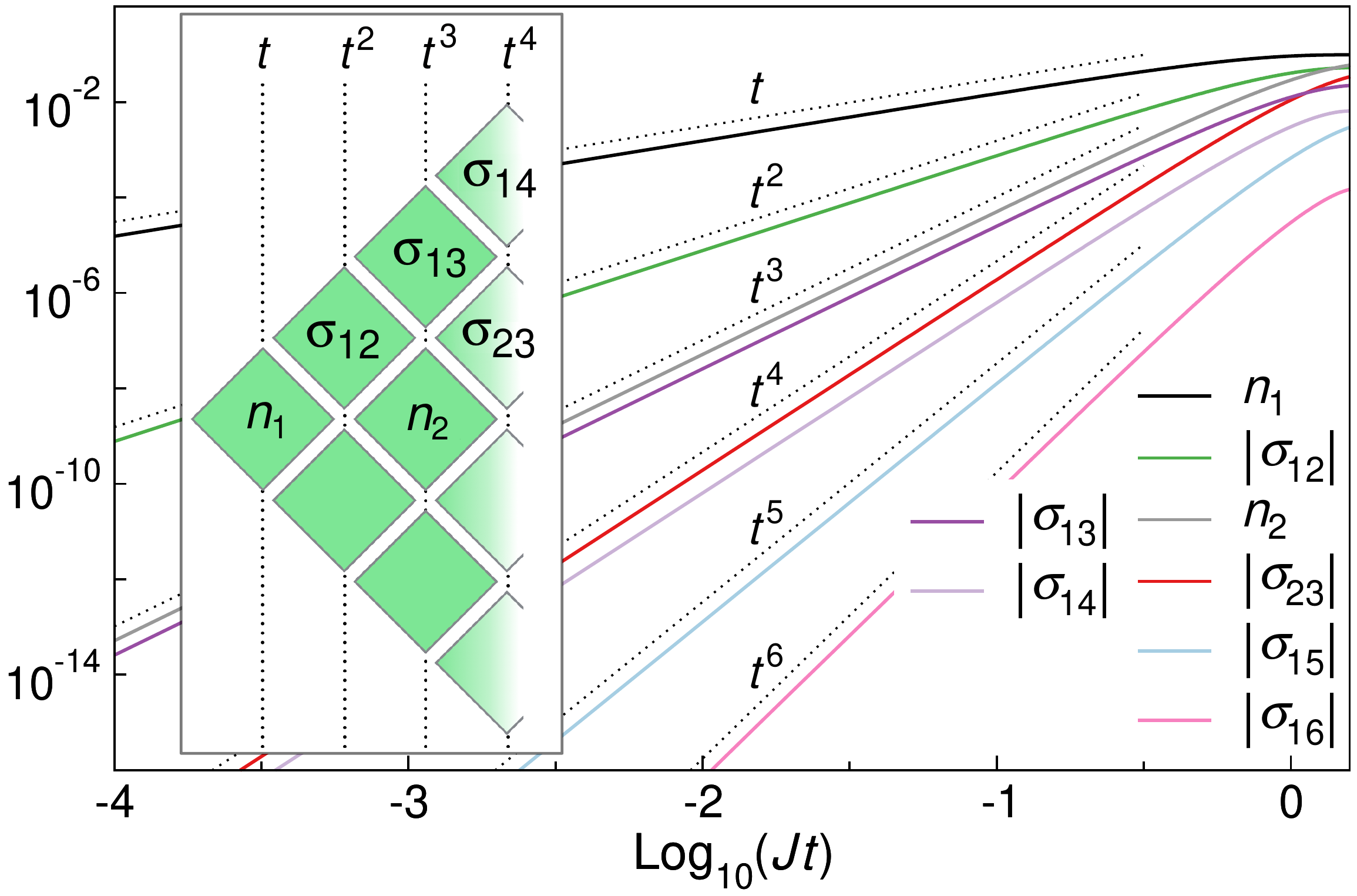}
\caption{Short time evolution of the single-particle density matrix components $\sigma_{jk}(t)$ [Eq.~\eqref{eq:defSPDM}] (solid lines) obtained numerically from Eqs.~\eqref{eq:bigSet}.
Dotted lines highlight the analytical power-law dependence predicted by Eq.~\eqref{eq:superSmallTDep}. 
The inset depicts the correlation between $\sigma_{jk}$ and the associated power-law exponent.
The parameters considered are the same as in Fig.~\ref{fig:TransFerFinRes}.}
\label{fig:smallTimes}
\end{figure}

The first stage of the dynamics is characterized by an increasing particle flow from the reservoirs into the system. Assuming that the the population redistribution in the reservoirs is negligible around $ t \approx 0 $, and hence taking $n_{L,R}(\varepsilon_S,t)\simeq n_{L,R}(\varepsilon_S,0)$, an iterative solution of Eqs.~\eqref{eq:bigSet} starting from $t=0$ provides the dominant time dependence of the SPDM components. As shown in Appendix \ref{app:SPDMsol}, in the general case one has 
\begin{equation}
|\sigma_{jk}(t \rightarrow 0)| \propto 
(J t)^{ M- |j+k -(M+1)|},
\label{eq:superSmallTDep}
\end{equation}
where the proportionality coefficient involves the term $\gamma_L n_L (\varepsilon_S,0)$ if $j+k<M+1$,  $\gamma_R n_R (\varepsilon_S,0)$ if $j+k>M+1$, and a combination of both if $j+k=M+1$.
As demonstrated in Fig.~\ref{fig:smallTimes}, the SPDM elements exhibit a power-law growth for short times, with exponents ranging from $1$ to $M$. 
From the perturbative solution elaborated in Appendix \ref{app:SPDMsol}, one obtains an estimate for the time scale that controls the validity of Eq.~\eqref{eq:superSmallTDep}, 
\begin{equation}
 \tau_0\lesssim \frac{1}{\gamma_{L,R}},
\end{equation}
in accord with intuition. 

After the initial loading of the lattice, the system dynamics are dominated by interferences of an increasing number of transmission and reflection amplitudes of single fermions tunneling across the lattice, leading to a rather strongly oscillating behaviour of single particle observables such as current and on-site populations. As observed on the bottom panel of Fig.~\ref{fig:TransFerFinRes}, it is on these time scales that also the SPDM long range coherences attain their maximum values. 

Given that $N_0\gg1$, 
the short time coherent dynamics ensuing from the time-dependent quantum-classical master equations \eqref{eq:coupledEquations} are consistent with those obtained for stationary reservoirs (cf.~Figs.~\ref{fig:exampleDynamics6} and \ref{fig:TransFerFinRes}). 
\subsubsection{Metastability and equilibration} 
As time grows, the interference induced oscillations of the SPDM elements are damped out, and a very slowly evolving state, which we refer to as \emph{metastable state}, emerges. The terminology used to describe this regime, although formally approximate, stems from the idea that the system is almost in a stable condition, which is progressively 
updated by the continuous variation of the reservoir bias. 
 This metastable state steadily converges towards a final equilibrium, on a rather long time scale.

As observed in Fig.~\ref{fig:TransFerFinRes}, the metastable state is characterized by quasi-stationary values of the SPDM elements with strongly suppressed long-range coherences. In particular, all the site-to-site currents are effectively the same, and the number $N_S(t)$ of particles on the lattice remains approximately constant, which in turn requires that $\dot{N}_L(t)+\dot{N}_R(t)\simeq0$, due to particle conservation.
Therefore, the emergence of metastability signals a balanced exchange of particles between the reservoirs and hence the emergence of a steady macroscopic current through the system, $I(t)$, which is
defined as \cite{Datta1995,Lan1957,But1986,Mei1992,Krinner2017,Leb2018}
\begin{equation}
I (t) := - \frac{1}{2} \frac{d}{dt } \Delta N (t),
\label{eq:bigCurrent}
\end{equation}
where $\Delta N(t)=N_L(t)-N_R(t)$. From the rate equations [Eqs.~\eqref{eq:NLevolution} and \eqref{eq:NRevolution}] for the reservoir particle numbers, one infers
\begin{equation}
 I(t) =\frac{-\gamma_L}{2} [ n_1 (t )- n_L (\varepsilon_{S}, t)] + \frac{\gamma_R}{2} [ n_M (t )+  n_R (\varepsilon_{S}, t)].
 \label{eq:bigCurrEqAbu}
\end{equation}
The connection between $I(t)$ and the local particle 
flow on the lattice is inferred from the evolution [Eqs.~\eqref{eq:coupledevolutionSPDM}] of the on-site number of particles, which implies 
\begin{align} 
 \dot{n}_1-\dot{n}_M & = -j_{12}(t) -j_{M-1,M}(t) + 2 I(t), \\
 \dot{n}_l &= -j_{l,l+1}(t) + j_{l-1,l}(t), \quad 2\leqslant l\leqslant M-1,
\end{align}
where the definition of the site-to-site current in Eq.~\eqref{eq:jcurrentdef} was used. In the metastable regime, the time variation of the on-site population is negligible, and the equations above imply the homogenization of all site-to-site currents, $j_{l,l+1}(t)= j(t)$, as well as, most importantly, the identity of macroscopic and local currents:
\begin{equation}
 I(t)=j(t).
 \label{eq:CurrentConsistency}
\end{equation}
We stress that this fundamental consistency of the transport process is always ensured by our formalism of classical-quantum master equations \eqref{eq:coupledEquations} once the metastable regime is reached. 

As observed in Fig.~\ref{fig:exampleDynamics6}, in the metastable regime, the values of the long-range coherences and the time derivative of the current
may further be neglected 
in the SPDM evolution equations, which leads to 
\begin{subequations}
\begin{align}
n_{1 }  (t ) & \approx  \bar{n} (t ) +  \frac{ 4 (\gamma_L - \gamma_R) J^2 + \gamma_L \gamma_R^2 +  \gamma_L^2 \gamma_R }{ 2(4 J^2 + \gamma_L \gamma_R ) (\gamma_L + \gamma_R )} \Delta n (t), 
\label{eq:n1_Metastable}\\
n_{m } (t ) & \approx   \bar{n} (t ) +  \frac{ 4 (\gamma_L - \gamma_R) J^2 + \gamma_L \gamma_R^2 -  \gamma_L^2 \gamma_R }{ 2(4 J^2 + \gamma_L \gamma_R ) (\gamma_L + \gamma_R )} \Delta n (t),
\label{eq:nmm_Metastable} \\
n_{M } (t ) & \approx   \bar{n} (t ) +  \frac{ 4 (\gamma_L - \gamma_R) J^2 - \gamma_L \gamma_R^2 -  \gamma_L^2 \gamma_R }{ 2(4 J^2 + \gamma_L \gamma_R ) (\gamma_L + \gamma_R )} \Delta n (t),  
\label{eq:nM_Metastable}
\end{align}
\label{eq:populationsMetastable}
\end{subequations}
and 
\begin{equation}
j(t ) \approx \frac{ 4 \gamma_L \gamma_R J^2 }{ (4 J^2 + \gamma_L \gamma_R ) (\gamma_L + \gamma_R )} \Delta n (t),
\label{eq:currentMetastable}
\end{equation} 
in terms of the time dependent populations of the reservoir resonant energy levels, 
\begin{align}
\Delta n (t) = & \textrm{ } n_L ( \varepsilon_S ,t) - n_R ( \varepsilon_S ,t), \label{eq:deltantime}\\
\bar{n} (t) = & \textrm{ } \frac{ n_L ( \varepsilon_S, t ) + n_R ( \varepsilon_S,t)}{2}.
\label{eq:barntime}
\end{align}
Obviously, these functional relations between the populations of resonant reservoir states, the local current and lattice occupation constitute a time-dependent generalization of the NESS for stationary reservoirs discussed in Sec.~\ref{sec:NumDynInfRes}. This comparison allows us to interpret the 
metastable many-particle state on the lattice as a continuously parametrized sequence of stationary states which are determined by the continuously updating reservoir states. 
Let us also emphasize that inserting Eqs.~\eqref{eq:n1_Metastable} and \eqref{eq:nM_Metastable} into Eq.~\eqref{eq:bigCurrEqAbu}, leads to the consistency condition \eqref{eq:CurrentConsistency}, as it must be.
 
The slow evolution of the metastable state will eventually come to an end. The evolution equations entail that equilibrium is achieved (currents vanish) once the populations of the resonant energy levels in the reservoirs and those of the lattice sites coalesce. Since the particle exchange between the reservoirs in the metastable regime is governed by $j(t)$, and thus proportional to $\Delta n(t)$, we can expect that equilibrium will not be achieved in a finite time, but rather asymptotically as $t\to\infty$.  The equilibrium state is then characterized by the population 
\begin{equation}
 n^\infty\equiv n_j(t\to\infty)=n_L(\varepsilon_S,t\to\infty)=n_R(\varepsilon_S,t\to\infty),
\end{equation}
which is related to the equilibrium value of the chemical potential $\mu^\infty$ through Eq.~\eqref{eq:occN}. Since we assume both reservoirs to be at the same temperature, the equilibrium state also has an equal final number of particles in the reservoirs, $N^\infty\equiv N_{L,R}(t\to\infty)$. Using particle number conservation, the equilibrium condition can be calculated from
\begin{equation}
 N_0 = 2N^\infty + M n^\infty,
 \label{eq:equilibrium_def}
\end{equation}
solving numerically for $\mu^\infty$. 

While Eqs.~\eqref{eq:populationsMetastable} and \eqref{eq:currentMetastable} provide the approximate SPDM dynamics in terms of the evolution of the reservoirs, the precise time dependence of $\Delta n(t)$ and $\bar{n}(t)$ when approaching equilibrium remains to be determined. For this purpose, we rewrite the evolution equations for the chemical potentials [Eqs.~\eqref{eq:evolution_muL} and \eqref{eq:evolution_muR}] in terms of the resonant populations,
\begin{subequations}
\begin{align}
 \frac{d}{dt} n_L (\varepsilon_S,t) &=   \dfrac{g(\mu_L(t), \beta, \varepsilon_S)}{f ( \mu_L (t) ,\beta,  E_0 )}  \gamma_L \left[  n_1 (t )- n_L (\varepsilon_{S}, t) \right], \label{eq:evolution_nL}\\
 \frac{d}{dt} n_R (\varepsilon_S,t) &=   \dfrac{g(\mu_R(t), \beta, \varepsilon_S)}{f ( \mu_R (t) , \beta,  E_0)}  \gamma_R \left[  n_M (t )- n_R (\varepsilon_{S}, t) \right]. \label{eq:evolution_nR}
\end{align}
\label{eq:evolution_nLnR}
\end{subequations}
To derive the first non-vanishing contribution of the asymptotic evolution, we evaluate the ratio of the functions $g$ and $f$ at $t=\infty$ and use the metastable form for $n_1(t)$ and $n_M(t)$ given in Eqs.~\eqref{eq:populationsMetastable}. 
Then, subtracting Eq.~\eqref{eq:evolution_nR} from Eq.~\eqref{eq:evolution_nL} leads to a differential equation for $\Delta n(t)$, 
\begin{equation}
 \frac{d}{dt} \Delta n(t)= -\alpha\, \Delta n(t),
\end{equation}
where 
\begin{equation}
  \alpha \equiv \dfrac{g(\mu^\infty,\beta,\varepsilon_S)}{f ( \mu^\infty,\beta,  E_0 )} \frac{ 8 \gamma_L \gamma_R J^2 }{ (4 J^2 + \gamma_L \gamma_R ) (\gamma_L + \gamma_R )}.
  \label{eq:defalpha}
\end{equation}
Therefore, the population gap between the resonant energy levels in the reservoirs closes exponentially as
\begin{equation}
 \Delta n(t\to\infty) \propto e^{-\alpha t}.
 \label{eq:smallndecay}
\end{equation}
\begin{figure}
\centering
\includegraphics[width=0.98\columnwidth]{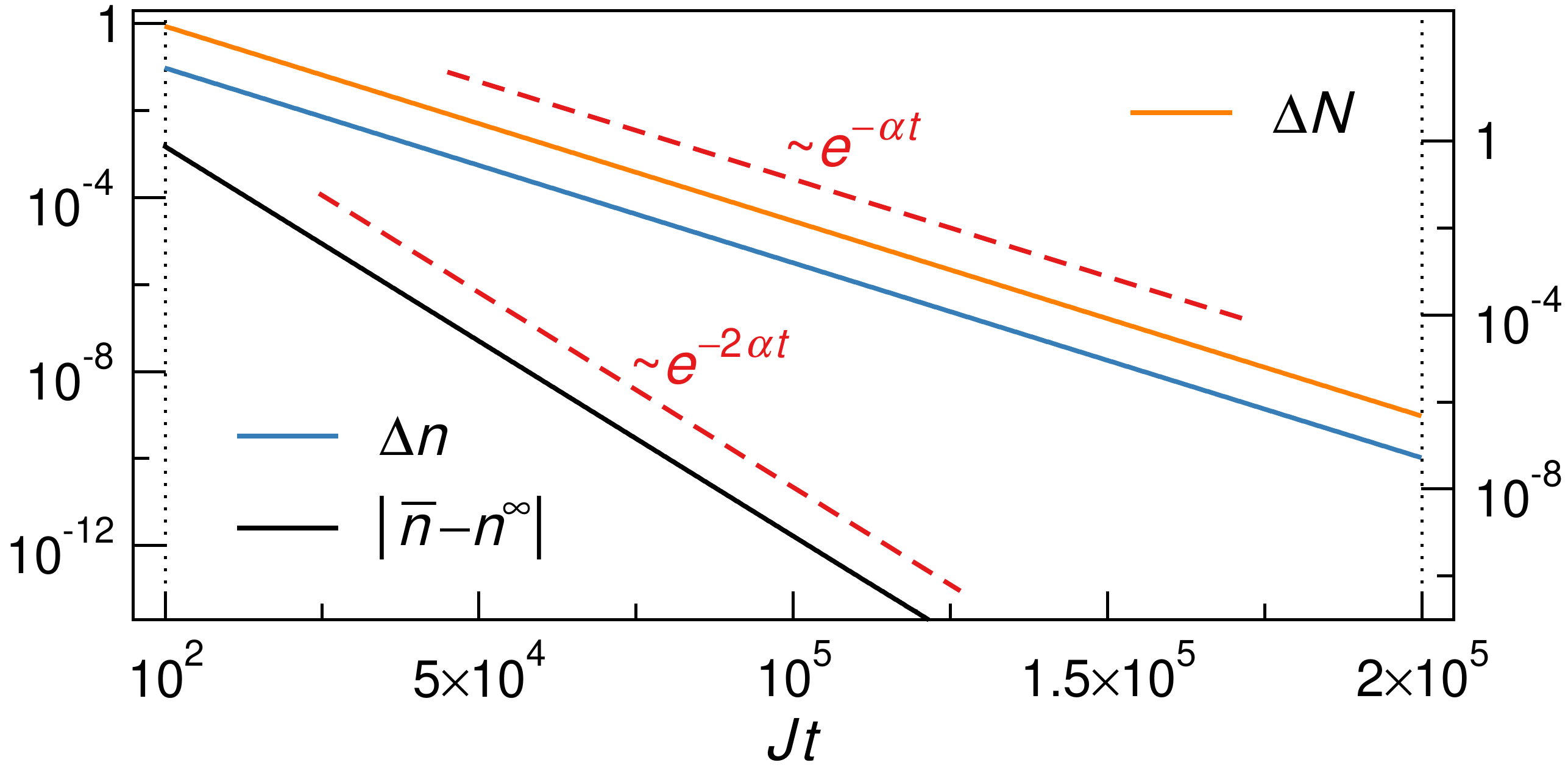}
\caption{Long-time decay of the reservoir particle number difference $\Delta N(t)$ (right ordinate), occupation number difference $\Delta n(t)$ [Eq.~\eqref{eq:deltantime}] and average $\bar{n}(t)$ [Eq.~\eqref{eq:barntime}] (left ordinate) with respect to their corresponding equilibrium values, for fermionic transport between finite (non-stationary) reservoirs. 
Dashed red lines indicate exponential decays with rates $ \alpha $ and $ 2 \alpha $, respectively, for $\Delta n(t)$ [Eq.~\eqref{eq:smallndecay}] and  $\bar{n}(t)$ [Eq.~\eqref{eq:smallnavdecay}], with $\alpha/J= 1.032\times 10^{-4}$ obtained according to Eq.~\eqref{eq:defalpha}, for the presently used parameter values (same as in Fig.~\ref{fig:TransFerFinRes}). 
}
\label{fig:LongTimeDecayFinRes_DeltaNn}
\end{figure}
\begin{figure*}
\centering
\includegraphics[width=0.95\textwidth]{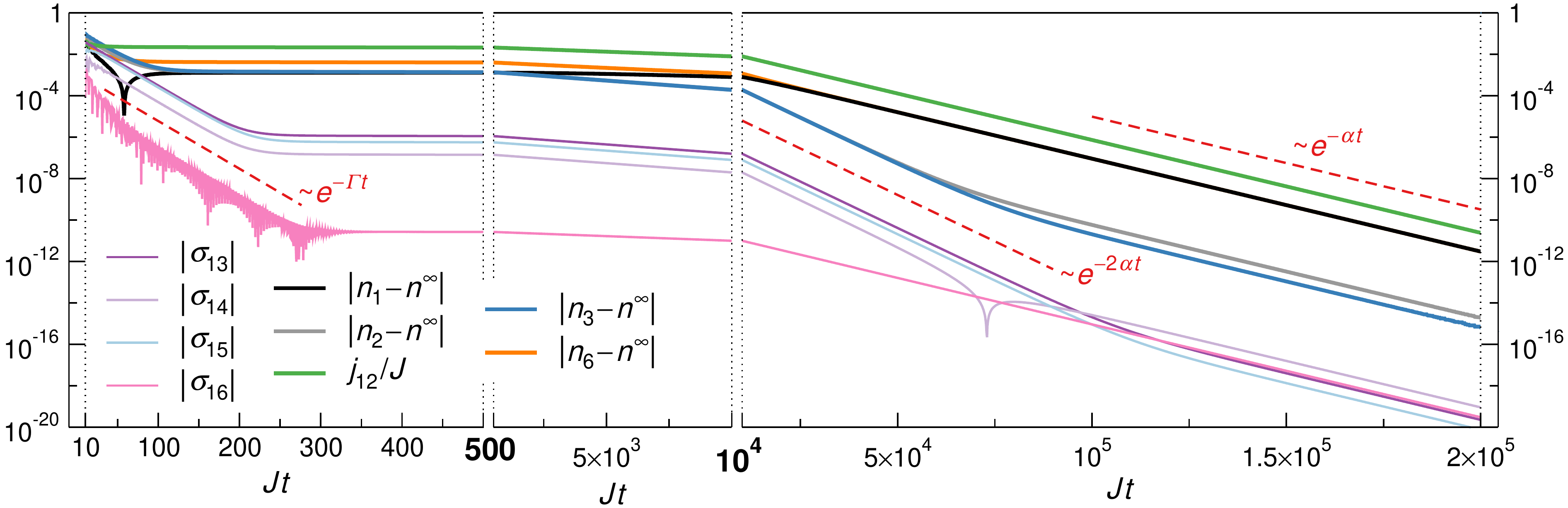}
\caption{Long-time dynamics of fermionic transport between finite (non-stationary) reservoirs: Decay of lattice populations $n_l(t)$, current $j_{12}(t)$, and long-range coherences $\{\sigma_{1j}\}_{j=3,\ldots,6}$ with respect to their corresponding equilibrium values.
Note the different scales on the time axis, changing at the values highlighted in bold. 
 Dashed red lines mark the indicated exponential decays for $\Gamma/J=0.0530209$ [obtained from the diagonalization of $H_\text{eff}$ in Eq.~\eqref{eq:effHam} and defining the relaxation time scale \eqref{eq:relaxTime}], and for $\alpha/J= 1.032\times 10^{-4}$ obtained from Eq.~\eqref{eq:defalpha}. The parameters used are the same as in Fig.~\ref{fig:TransFerFinRes}.} 
\label{fig:LongTimeDecayFinRes}
\end{figure*}

Similarly, we obtain a differential equation for $\bar{n}(t)$. In this case, a first order expansion of the $g/f$ ratio around $\mu^\infty$ is necessary to obtain the first non-vanishing contribution to the evolution. 
Then, by adding Eqs.~\eqref{eq:evolution_nLnR}, we get 
\begin{equation}
 \frac{d}{dt} \bar{n}(t) \propto [\Delta n(t)]^2,
\end{equation}
which yields the asymptotic dependence 
\begin{equation}
 |\bar{n}(t\to\infty)-n^\infty| \propto e^{-2\alpha t}.
 \label{eq:smallnavdecay}
\end{equation}
The predictions [Eqs.~\eqref{eq:smallndecay} and \eqref{eq:smallnavdecay}] for the asymptotic behaviour of $\Delta n(t)$ and $\bar{n}(t)$ are in perfect agreement with the numerical simulations, as apparent from Fig.~\ref{fig:LongTimeDecayFinRes_DeltaNn}. 

The time dependence of $\Delta n(t)$ carries over to the currents $j(t)$ and $I(t)$, as well as to $\Delta N(t)$. Hence, our formalism predicts an exponentially decreasing macroscopic current between finite reservoirs, independently of the specific system parameters and of the bosonic or fermionic nature of the particles. Such exponentially decreasing current between equilibrating reservoirs has been experimentally observed in cold atom experiments \cite{Krinner2017}. 

In Fig.~\ref{fig:LongTimeDecayFinRes} we provide a complete picture of the time evolution of the SPDM elements, over several orders of magnitude on the time axis, after the short-time coherent regime. 
We initially observe the relaxation time scale as predicted by Eq.~\eqref{eq:relaxTime} for stationary reservoirs, but eventually the exponential rate decreases from $ \tau_{\textrm{rel}}^{-1}$ to $ \alpha$. This is a consequence of the correlation of the dynamics on the lattice and in the reservoirs, due to the finite total particle number. 
We note that the on-site populations, which in this regime fulfil Eqs.~\eqref{eq:populationsMetastable}, can exhibit both the $ \alpha $ and the $2 \alpha $ decay rates [as observed for $ n_2 (t) $ and $ n_3 (t) $ in Fig.~\ref{fig:LongTimeDecayFinRes}], since they are linear combinations of $ \Delta n (t) $ and $ \bar{n} (t) $, which obey Eqs.~\eqref{eq:smallndecay} and \eqref{eq:smallnavdecay}, respectively. 
Both rates may also appear in the decay of long-range coherences, whereas  
the dependence of the site-to-site current on $\Delta n (t)$, given in Eq.~\eqref{eq:currentMetastable}, leads to its unique long-time decay rate $\alpha$ (cf.~Fig.~\ref{fig:LongTimeDecayFinRes}). 

We have thus demonstrated that the finite size of reservoirs induces a new (longer) dynamical equilibration time scale, 
\begin{equation}
 \tau_{\textrm{eq}} := \frac{1}{\alpha},
 \label{eq:def_tauE}
\end{equation}
which, by virtue of \eqref{eq:defalpha}, depends, in particular, on the effective coupling strengths $\gamma_{L,R}$ 
between the lattice's terminal sites and the adjacent reservoirs, as well as on the initial state of the reservoirs (which enters Eq.~\eqref{eq:defalpha} via $\mu^\infty$ 
in the functions $f$ and $g$ [Eqs.~\eqref{eq:def_f} and \eqref{eq:def_g}]).
Whereas $\mu^\infty$ slightly changes with the lattice length [recall Eq.~\eqref{eq:equilibrium_def}], 
the equilibration time scale is effectively system size independent for $M\ll N_0$. 
 
Furthermore, we obtain a simple approximation for $ \alpha $, in terms of the reservoir initial conditions and the system physical parameters, noticing that the reservoir particle number difference $ \Delta N (t) $, and the resonant energy level occupation number difference $ \Delta n (t) $ decay with the same rate in the long-time limit [recall Eq.~\eqref{eq:smallndecay} and Fig.~\ref{fig:LongTimeDecayFinRes_DeltaNn}]. 
Substituting Eq.~\eqref{eq:currentMetastable} in Eq.~\eqref{eq:bigCurrent}, via Eq.~\eqref{eq:CurrentConsistency}, we obtain the differential equation
\begin{equation}
\frac{d}{dt } \Delta N (t)= - \frac{ 8 \gamma_L \gamma_R J^2 }{ (4 J^2 + \gamma_L \gamma_R ) ( \gamma_L + \gamma_R ) } \Delta n (t) ,
 \label{eq:bigNsmallN}
\end{equation}
which, integrated
 from a certain time $t^*$ (which can be identified with the onset of metastability) to $t=\infty$, 
  leads to
\begin{equation}
\Delta N (t^*) = \frac{1 }{ \alpha }  \frac{ 8  \gamma_L \gamma_R J^2 }{ (4 J^2 + \gamma_L \gamma_R ) ( \gamma_L + \gamma_R ) } \Delta n (t^*) ,
 \label{eq:alphaEstimation}
\end{equation}
since $ \Delta N (\infty) = \Delta n (\infty) = 0 $. 
For $N_0\gg 1$, the reservoirs do not evolve noticeably during the coherent dynamics regime before the onset of metastability at $t^*$, 
hence we can roughly approximate 
$\Delta n (t^*) / \Delta N (t^*) \approx \Delta n (0) / \Delta N (0) $, which 
leads to 
\begin{equation}
\alpha \approx \frac{ 8  \gamma_L \gamma_R J^2 }{ (4 J^2 + \gamma_L \gamma_R ) ( \gamma_L + \gamma_R ) } \frac{ \Delta n (0)  }{ \Delta N (0)}  .
 \label{eq:alphaEstimationFinal}
\end{equation}

\section{Two-particle density matrix: Bosonic versus fermionic currents} 
\label{sec:tpdm}
We stress that the treatment presented above is valid for both bosonic and fermionic particles, and, thus, well suited to explore which features 
in the transport phenomena reveal the nature of the carriers.  
One obvious difference in the dynamics is seeded by Pauli's exclusion principle, ensuring $n_{L,R}(\varepsilon_S,t)\leqslant 1$ and $n_l(t)\leqslant 1$ at all times for fermionic transport. Let us, however, consider the case in which 
 the chosen initial conditions guarantee that both limitations are respected, for bosons as well as for fermions, e.g. by chosing  $n_{L,R}(\varepsilon_S,0)<1$ and $n_l(0)=0$.

Recall that the SPDM equations \eqref{eq:bigSet}, while looking formally identical for bosons and fermions, encode distinct nonlinearities in the function $f(\mu(t),\beta,E_0)$ determined by the reservoir particle distribution. In the time interval where $S$ exhibits coherent oscillations, and in the regime of applicability of our formalism ($N_0\gg 1$), the environment change is negligible and thus there will be no difference between the SPDM evolutions of both types of carriers. Therefore, for short time dynamics, one cannot discriminate between bosonic and fermionic transport by inspection of single particle system observables. This is consistent with the fact that, for non-interacting particles, the evolution of one-particle observables is insensitive to many-particle interference effects \cite{Mayer2011}. Differences, nonetheless, could arise in the short time evolution of the SPDM when including quartic terms in the system Hamiltonian \eqref{eq:lattHamiltonian}, i.e., for interacting particles \cite{Bru2018}. 

Interestingly, the dynamics of higher-order observables, such as the two-particle density matrix (TPDM), 
\begin{equation}
\Delta_{jmkn} (t) = \Tr_S [ a_j^{\dagger} a_m a_k^{\dagger} a_n \rho_S (t) ],
\end{equation}
we expected to expose many-particle interference effects for non-interacting carriers, and hence to reveal clear-cut differences between bosonic and fermionic transport.
From Eq.~\eqref{eq:meLindbladtime}, the evolution equations for  $\Delta_{jmkn} (t)$ read 
\begin{widetext}
\begin{align}
\dot{\Delta}_{jmkn} =&  \textrm{ } i  J [ \Delta_{j,m+1,kn} + \Delta_{j,m-1,kn} + \Delta_{jmk,n+1} + \Delta_{jmk,n-1} 
 - \Delta_{j+1,mkn} - \Delta_{j-1,mkn} - \Delta_{jm,k+1,n} - \Delta_{jm,k-1,n} ] \notag\\
 &- \frac{ \gamma_L }{2} [\delta_{j1} +  \delta_{k1} + \delta_{1m} +  \delta_{1n}] \Delta_{jmkn} 
 + \gamma_L \delta_{m1}\delta_{k1} \sigma_{jn} \notag \\
 &+ \gamma_L n_L(\varepsilon_S,t) 
 \left[ \delta_{k1} \delta_{n1} \sigma_{jm} + \delta_{j1} \delta_{m1} \sigma_{kn} + \delta_{j1}\delta_{n1} \left(\delta_{km} \pm \sigma_{km} \right) \pm \delta_{m1}\delta_{k1} \sigma_{jn}\right] \notag \\
 &+ \textrm{ } ( \{1, L\} \leftrightarrow \{M,R\} ), 
\label{eq:tpdmEquations}
\end{align}
\end{widetext}
which are given in terms of the known evolution of the SPDM [$\sigma_{jk}(t)$] and of the reservoir resonant level populations $n_{L,R}(\varepsilon_S,t)$.
Equations \eqref{eq:tpdmEquations} distinctly depend on the nature of the particles (upper sign $+$ for bosons, lower sign $-$ for fermions). 
Among all two-particle observables, the fluctuations of single particle observables, such as the on-site particle number and site-to-site currents, are most accessible and can be obtained from the elements of the TPDM,
\begin{align}
 \Delta^2n_l(t) \equiv &\; \Delta_{llll}-n_l^2(t), \\
 \Delta^2j_{l,l+1} \equiv &\; J^2\big(\Delta_{l+1,l,l,l+1}+\Delta_{l,l+1,l+1,l} \notag \\
 &- \Delta_{l+1,l,l+1,l} -\Delta_{l,l+1,l,l+1}\big) - j_{l,l+1}^2(t).
\end{align}
\begin{figure}
\centering
\includegraphics[width=0.95\columnwidth]{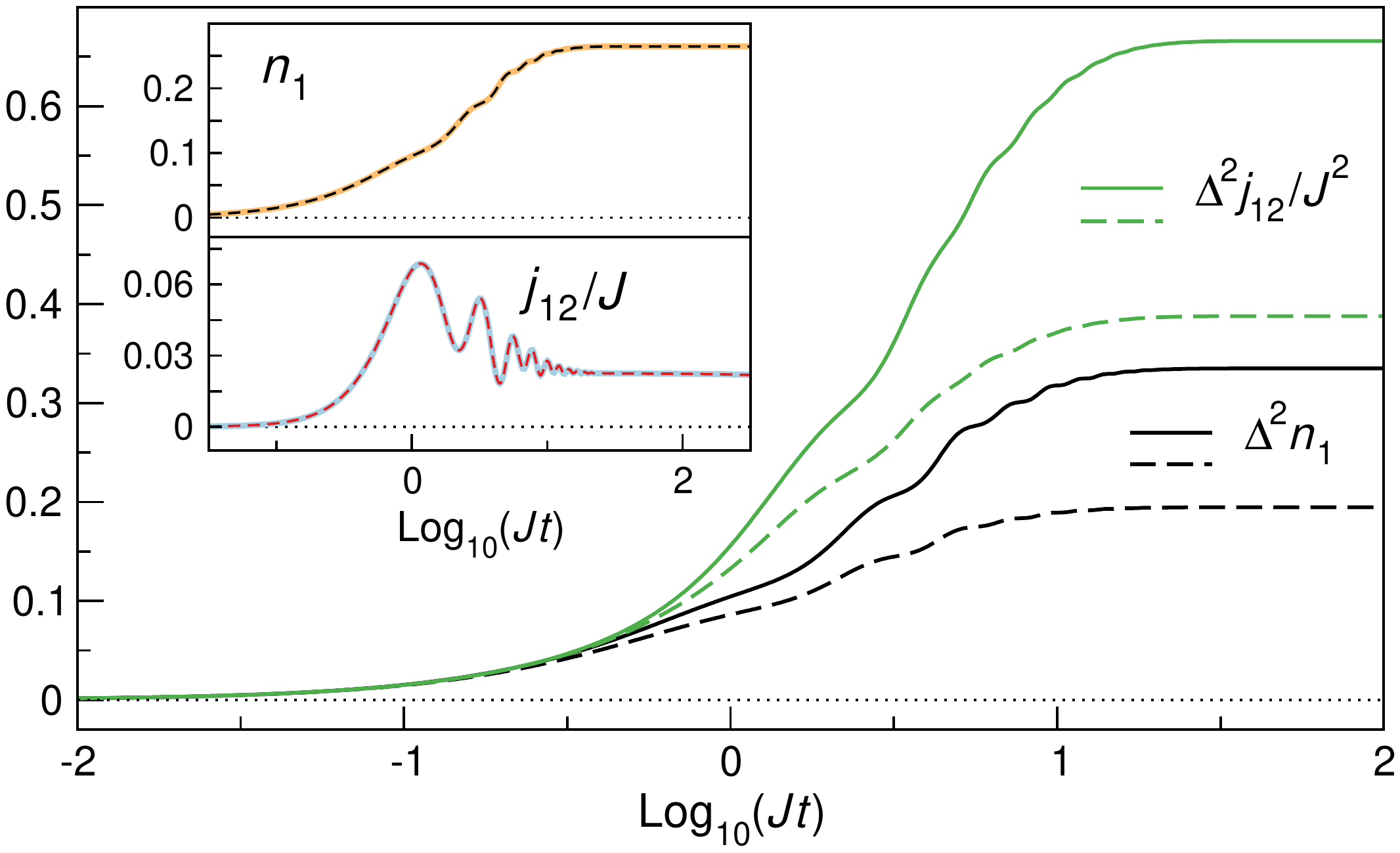}
\caption{Signatures of many-particle interference  in quantum transport  between finite (non-stationary) reservoirs: Time evolution of the on-site population variance $\Delta^2 n_1(t)$ and of the current variance $\Delta^2 j_{12}(t)$ for bosons (solid lines) and fermions (dashed lines). 
The inset shows the evolution of the corresponding single particle observables, $n_1(t)$ and $j_{12}(t)$. 
The reservoir configuration for fermions is  $\beta^{(F)}J=1$, $\mu_L^{(F)}(0)=1.2J$ [$N_L^{(F)}(0)=1276$], $\mu_R^{(F)}(0)=0.7J$ [$N_R^{(F)}(0)=838$], and for bosons $\beta^{(B)}J=0.7$, 
$\mu_L^{(B)}(0)=-0.059J$ [$N_L^{(B)}(0)=1654$] , $\mu_R^{(B)}(0)=-0.479J$ [$N_R^{(B)}(0)=1164$]. The remaining parameter values are, in all cases, $n_L (\varepsilon_S,0) =0.310$, $n_R (\varepsilon_S,0) =0.214$, $\gamma_L = \gamma_R = 0.5 J$, $\varepsilon_S = 2J$, $M=3$.}
\label{fig:tpdm}
\end{figure}

In Fig.~\ref{fig:tpdm}, we show the dynamics of these variances and confirm the distinct behaviour of bosonic and fermionic particles. 
Note in particular that, due to the fermionic anticommutation relations, the fluctuation of the fermionic on-site populations is bounded from above, $\Delta^2n_l(t)=n_l(t)[1-n_l(t)]\leqslant 1/4$; a limitation that does not apply for bosons. 
 For the initially empty lattice here considered, the fluctuations become most dissimilar, and thus the fermionic or bosonic character is most recognizable, once the metastable state is reached. 

\section{Conclusions}
We have conceived a novel formalism to study many-particle quantum transport across a system locally coupled to two finite ---non-stationary--- reservoirs. This approach goes beyond the standard open quantum system treatment where the environment is assumed to be stationary in time, and which can only account for the emergence of non-equilibrium steady states with non-vanishing currents. 
We showed that a set of coupled (nonlinear) quantum-classical master equations can describe the correlated dynamics of a system and of two finite size (bosonic or fermionic) particle reservoirs, which evolve through grand canonical thermal states characterized by time-dependent chemical potentials and a common temperature. This construction rests on the assumption of validity of the local master equation [Eqs.~\eqref{eq:coupledEquations}] and ensures the conservation of the total particle number. 

We have shown that the coherent short time dynamics on a uniform one-dimensional lattice are characterized by a power-law growth of the single particle density matrix elements, followed by a regime dominated by single particle interference, during which the 
variation of the reservoirs is negligible. The coupling to the reservoirs first manifests itself as an exponential relaxation, characterized by the time scale $\tau_{\textrm{rel}}$ [Eq.~\eqref{eq:relaxTime}], 
of the coherent system dynamics. 
The finiteness of the reservoirs, however, soon leads to a change of dynamical regime and 
the emergence of a metastable state, which is characterized by a slowly varying macroscopic current between the reservoirs. This current decreases exponentially in time, independently of the specific system parameters and of the nature of the particles, and is consistent with the internal site-to-site currents. The slow evolution of this metastable state populates a final equilibrium state with a vanishing particle flow and homogeneous population of lattice and resonant reservoir energy levels. The exponential convergence towards equilibrium is governed by a new time scale $\tau_{\textrm{eq}}$ [Eq.~\eqref{eq:def_tauE}], whose dependence on all system and reservoir parameters is analitically given. 

Our approach is well suited to investigate many-particle interference effects on quantum transport, as here illustrated by discriminating bosonic and fermionic carriers by inspection of the fluctuations of currents and on-site populations. 

Let us finally emphasize that our formalism, treatment and results, 
apply to general scenarios beyond the uniform lattice model coupled to 3D harmonic traps here chosen to numerically support our findings.
\begin{acknowledgments}
The authors are grateful to the German Research Foundation (DFG project WI3426/7-1), as well as to the state of Baden-W\"urttemberg through bwHPC, and thank Martin Lebrat for helpful discussions. 
GA furthermore acknowledges support by Fondazione Grazioli and by DAAD. 
\end{acknowledgments}
\appendix
\section{3D harmonic trap reservoirs}
\label{sec:AppHarTrap}
We consider here the specific case of particle reservoirs described by 3D harmonic traps, with frequency $ \omega_i $ along direction $i$. The reservoir density of states in the continuum limit can be analytically computed and reads
\begin{equation}
 D (\varepsilon ) = \frac{\varepsilon^2}{ 2 \hbar^3 \omega_x \omega_y \omega_z  }. 
\end{equation}

Knowing that the minimum energy $E_0$ of the reservoirs is easily expressed in terms of the trapping frequencies,
\begin{equation}
 E_0 = \frac{\hbar}{2} ( \omega_x + \omega_y + \omega_z ),
\end{equation}
one can obtain a close expression for the reservoir particle number, 
\begin{align}
 N_L(t)  =& \frac{1}{\hbar^3 \omega_x \omega_y \omega_z} \bigg\{
 \mp\frac{E_0^2}{2\beta} \log\left[1 \mp e^{-\beta[E_0-\mu_L(t)]}\right] \notag \\
 &\pm \frac{E_0}{\beta^2}\Li_2 \left(\pm e^{ - \beta [ E_0 - \mu_L (t)] } \right)  \notag \\ 
 &\pm \frac{1}{\beta^3}\Li_3 \left(\pm e^{ - \beta [ E_0 - \mu_L (t)] } \right) \bigg\},
\end{align}
and for the function connecting $\dot{N}_L(t)$  
and the chemical potential [Eq.~\eqref{eq:def_f}], 
\begin{align}
 f ( \mu_L (t), \beta, E_0 )  =& \frac{1}{\hbar^3 \omega_x \omega_y \omega_z} \bigg\{
 \frac{E_0^2}{2} \frac{e^{-\beta[E_0-\mu_L(t)]}}{1\mp e^{-\beta[E_0-\mu_L(t)]}} \notag \\
 & \mp \frac{E_0}{\beta} \log\left[1 \mp e^{-\beta[E_0-\mu_L(t)]}\right] \notag \\
 &\pm \frac{1}{\beta^2}\Li_2 \left(\pm e^{ - \beta [ E_0 - \mu_L (t)] } \right)  \bigg\}
\end{align}
where, in both expressions, the upper (lower) choice of signs corresponds to the bosonic (fermionic) case. 
Note that for bosons (fermions) it must be $\mu_L(t)< E_0$ $(\mu_L(t)> E_0)$.
The expressions for the right reservoir are formally the same. 
\section{Short time solution of SPDM's equations}
\label{app:SPDMsol}
Let us consider the equations of motion for the SPDM [Eq.~\eqref{eq:bigSet}] under the approximation $n_{L,R}(\varepsilon_S,t)\simeq n_{L,R}(\varepsilon_S,0)$, i.e., for short times. 
The equations for the first few components read explicitly 
\begin{subequations}
\begin{align}
\frac{d}{dt} {\sigma}_{11}  = & \gamma_L [ n_L (\varepsilon_S,0) - {\sigma}_{11} ]  + i J [ \sigma_{12} - \sigma_{21}] , \\
\frac{d}{dt} {\sigma}_{12}  = & - \frac{ \gamma_L }{2} {\sigma}_{12} + i J [ \sigma_{11} +  \sigma_{13} - \sigma_{22}] , \\
\frac{d}{dt} {\sigma}_{13}  = & - \frac{ \gamma_L }{2} {\sigma}_{13} + i J [ \sigma_{12} - \sigma_{23} - \sigma_{14}]  , \\
 \vdots & \notag \\
\frac{d}{dt} {\sigma}_{22}  = & i J [ \sigma_{21} + \sigma_{23} - \sigma_{12} - \sigma_{32}] ,  \\
 \vdots & \notag
\end{align}
\label{eq:SPDMappApp}
\end{subequations}
Note that 
the SPDM is by construction a hermitian matrix, i.e., it is enough to consider $\sigma_{jk}(t)$ for $k\geqslant j$. Also, 
due to the structure of the equations 
and the choice of the vacuum state as initial condition for the lattice, 
the components of the SPDM 
are real for even $j+k$, and purely imaginary for odd $j+k$.

We proceed by solving Eqs.~\eqref{eq:SPDMappApp} iteratively from $t=0$ for $j+k<M+1$, i.e., for those SPDM elements which are  `closer' in time to the left reservoir. 
After substituting on the right hand side of the equations the initial values $\sigma_{jk}(0)=0$, 
the resulting system has a non trivial equation only for $ {\sigma}_{11} (t) $, 
 and yields the solution 
\begin{subequations}
\begin{align}
 \sigma_{11}^{(1)} (t)  &= \gamma_L n_L (\varepsilon_S,0 ) t, 
\end{align}
\label{eq:SPDMstep1}
\end{subequations}
and $\sigma_{jk}^{(1)} (t) = 0$ for any other element with $j+k<M+1$. The superindex ${}^{(1)}$ indicates that this is the solution up to first order in $t$, around $t=0$.

After inserting this solution back into the right hand side of Eqs.~\eqref{eq:SPDMappApp}, the SPDM elements up to order  $t^2$  are obtained, 
\begin{subequations}
\begin{align}
  \sigma_{11}^{(2)} (t)  &= \gamma_L n_L (\varepsilon_S,0 ) \left( t - \gamma_L t^2/2 \right), \\
  \sigma_{12}^{(2)} (t)  &= i J \gamma_L n_L (\varepsilon_S,0) t^2/2 , 
\end{align}
\end{subequations}
and $\sigma_{jk}^{(2)} (t) = 0$ for any other element with $j+k<M+1$. 
Analogously, one can compute the next order correction, 
\begin{subequations}
\begin{align}
  \sigma_{11}^{(3)} (t)  &= \gamma_L n_L (\varepsilon_S,0) \left[ t - \gamma_L \frac{t^2}{2} + \frac{\gamma_L^2 - 2 J^2}{ 6} t^3 \right] , \\
  \sigma_{12}^{(3)} (t)  &= \frac{i}{2} J \gamma_L n_L (\varepsilon_S,0) \left( t^2 - \frac{\gamma_L}{2} t^3 \right) ,\\
  \sigma_{13}^{(3)} (t)  &= - \gamma_L n_L (\varepsilon_S,0)  J^2 t^3/6, \\
  \sigma_{22}^{(3)} (t)  &= \gamma_L n_L (\varepsilon_S,0 )  J^2 t^3/3, 
\end{align}
\label{eq:step2spdm}
\end{subequations}
and $\sigma_{jk}^{(3)} (t) = 0$ for any other element with $j+k<M+1$.
One can see that for $j+k<M+1$, the leading terms behave as $\sigma_{jk}(t)\propto (Jt)^{j+k-1}$. 

For the elements with $j+k>M+1$, which are `closer' in time to the right reservoir, the expressions are analogous to the ones above [with $\gamma_R$ and $n_R(\varepsilon_S,0)$ instead] and the leading terms behave as $\sigma_{jk}(t)\propto (Jt)^{(M+1-j)+(M+1-k)-1}$. One thus arrives at the expression given in Eq.~\eqref{eq:superSmallTDep}.
%
\end{document}